\documentclass[aps,10pt,twocolumn,superscriptaddress,balancelastpage,showpacs,letterpaper,nofootinbib]{revtex4-2}

\usepackage[utf8]{inputenc}
\usepackage[T1]{fontenc}

\makeatletter
\AtBeginDocument{\let\LS@rot\@undefined}
\makeatother

\usepackage{amsmath,bbm,amssymb,bm} 
\usepackage{physics, mathtools, accents}

\usepackage[bbgreekl]{mathbbol}
\DeclareSymbolFontAlphabet{\mathbbm}{bbold}
\DeclareSymbolFontAlphabet{\mathbb}{AMSb}

\usepackage{nicematrix}
\usepackage{graphicx}

\usepackage[caption=false]{subfig} 

\usepackage{tikz}
\usetikzlibrary{arrows.meta, fit}
\usepackage{tabularx,multirow,booktabs}
\usepackage[normalem]{ulem}
\usepackage{xcolor}
\usepackage{comment}

\definecolor{darkblue}{rgb}{0,0,0.6}
\definecolor{darkred}{rgb}{0.6,0,0}
\definecolor{veludo}{RGB}{9,97,72} 
\definecolor{botan}{RGB}{193,50,142}
\definecolor{ruri}{RGB}{0,92,175}
\definecolor{LightGray}{gray}{0.9}

\usepackage{bibunits}
\usepackage[colorlinks=true,urlcolor=ruri,citecolor=ruri,linkcolor=botan]{hyperref}
\hypersetup{hypertexnames=false}
\usepackage{orcidlink}
\usepackage{soul}

\usepackage{etoolbox}
\usepackage{titletoc}
\definecolor{kaba}{RGB}{193,105,60}
\newcommand{\diag}{\operatorname{diag}}
\newcommand{\spec}{\operatorname{spec}}
\newcommand{\RePart}{\operatorname{Re}}
\titlecontents{section}[2.8em]
  {\addvspace{2pt}}
  {\contentslabel{2.6em}}
  {\hspace*{-2.6em}}
  {\titlerule*[0.6pc]{.}\contentspage}

\newcommand{\editcolor}{black}
\newcommand{\deleted}[1]{}

\newcommand{\ts}{\nobreak\hspace{.16667em}}

\tikzset{highlight/.style={rectangle, fill=red!15, rounded corners = 0.5 mm, inner sep=1pt, fit=#1}}

\begin{document}
\title{Neuromodulation--inspired gated associative memory networks: \\extended memory retrieval and emergent multistability}

\author{Daiki Goto\kern0.15em\orcidlink{0009-0004-5096-1341}$^\S$}
\email{dgoto@uchicago.edu}
\affiliation{Department of Physics, The University of Chicago, Chicago, Illinois 60637, USA}
\affiliation{The James Franck Institute, The University of Chicago, Chicago, Illinois 60637, USA}
\altaffiliation{These authors contributed equally to this work.}

\author{Hector Manuel Lopez Rios\kern0.15em\orcidlink{0000-0002-4276-7432}$^\S$}
\email{hectorlr@uchicago.edu}
\affiliation{The James Franck Institute, The University of Chicago, Chicago, Illinois 60637, USA}
\affiliation{Department of Chemistry, The University of Chicago, Chicago, Illinois 60637, USA}

\author{Monika Scholz\kern0.15em\orcidlink{0000-0003-2186-410X}}
\affiliation{Max Planck Institute for Neurobiology of Behavior—caesar, Bonn 53175, Germany}

\author{Suriyanarayanan Vaikuntanathan\kern0.15em\orcidlink{0000-0003-2431-6045}}
\email{svaikunt@uchicago.edu}
\affiliation{The James Franck Institute, The University of Chicago, Chicago, Illinois 60637, USA}
\affiliation{Department of Chemistry, The University of Chicago, Chicago, Illinois 60637, USA}

\begin{abstract} 

    \textbf{Abstract.} Classical autoassociative memory models have been central to understanding emergent computations in recurrent neural circuits across diverse biological contexts. 
    However, they typically neglect neuromodulatory agents that are known to strongly shape memory capacity and stability. 
    Here we introduce a minimal, biophysically motivated associative memory network in which neuromodulator-like signals are modeled by a self-adaptive, activity-dependent gating mechanism.
    Using many-body simulations and dynamical mean-field theory, we show that such gating fundamentally reorganizes the attractor structure: the network bypasses the classical spin-glass transition, maintaining robust, high-overlap retrieval far beyond the standard critical capacity\deleted{, without shrinking basins of attraction}. 
    Mechanistically, {\color{\editcolor} when the gates act as fully open or fully closed switches,} the gate stabilizes transient “ghost” remnants of stored patterns even far above the Hopfield limit, converting a single memory fixed point attractor into a cluster of fixed points. {\color{\editcolor} The retrieved state is set by the initial overlap rather than quantized to the stored pattern. This graded dependence may be exploited for input-driven computation with persistent activity}.
    These results establish that neuromodulation-like gating alone — without modifying synaptic weights or introducing explicit higher-order couplings — can dramatically expand memory capacity, eliminate catastrophic breakdown, and reshape a discrete attractor landscape into a rich continuum, suggesting a general and biologically accessible route to enhanced computational capabilities in neural circuits and neuromorphic architectures.
    \par\vspace{1\baselineskip}
    
    \textbf{Significance statement.} Neuromodulators such as neuropeptides and bioamines regulate behavioral states across species, yet their collective computational consequences remain poorly understood theoretically. We show that a single biologically motivated mechanism for neuromodulation, activity-dependent multiplicative gating of neuronal dynamics, is sufficient to overcome fundamental limitations of classical associative memory: it eliminates the catastrophic capacity ceiling, dramatically extends memory \deleted{capacity}{\color{\editcolor}retrieval} far beyond the Hopfield limit\deleted{, and preserves broad basins of attraction}. Gating further converts each discrete memory attractor into a continuous cluster of fixed points{\color{\editcolor}, so that the retrieved state varies with the initial overlap rather than being quantized to the stored pattern --- a feature {\color{\editcolor}useful} for input-driven computation with persistent activity}. These results suggest that neuromodulatory gating may provide richer computational capabilities in biological and neuromorphic architectures.
    
\end{abstract}

\maketitle



\section{Introduction}
Recent large-scale measurements of neuronal activity show remarkable collective population dynamics with large populations of neurons cycling repeatedly through brain states~\cite{perich2025}. These dynamics reflect internal states of the brain, but can also map to behavioral states~\cite{kato2015global,hallinen21,musall2019single,stringer2019spontaneous}. How the architecture of the brain sets up such widely coordinated patterns is an area of active research. Specifically, theoretical models with simple connectivity rules have shown the emergence of such patterns. For example, classical autoassociative memory models---exemplified by the Amari–Hopfield recurrent neural networks (RNNs)~\cite{Amari72,Hopfield82,Hopfield84}---embed a finite set of patterns directly in the interaction matrix so that the dynamics relax, given partial cues, to fixed-point attractors representing those patterns. 

Although simple, these classical autoassociative models have been used to model aspects of neuronal dynamics in living neuronal systems such as the hippocampus\ts\cite{mcnaughton1987hippocampal,treves1994computational} and cortex\ts\cite{amit1997model}. These models incorporate the pairwise coupling{\color{\editcolor}s} between neurons that mimic biological synapses, but do not incorporate potentially important extrasynaptic signaling contributions. Indeed, extrasynaptic activity due to neuromodulators (biogenic amines or neuropeptides) is ubiquitous and can influence the activity of large neuron populations, often via multiple parallel biochemical pathways \cite{nadim2014neuromodulation,jekely2024nonsynaptic}. For example, a recently completed identification of the neuropeptidergic connectome of the nematode \emph{Caenorhabditis elegans}~\cite{bentley2016multilayer,ripoll_neuropep_celegans,celegans_neuropep_rev}, revealed that the ``hard-wired'' synaptic network is overlaid by a complex ``wireless'' neuropeptidergic network. Beyond nematodes, these signaling modes are increasingly appreciated across other species, including in mice~\cite{zhou2024neuropeptides} and humans~\cite{bazzari2019neuromodulators}. Broad perturbations of neuromodulators affect behavioral state and state transitions \cite{marder2012neuromodulation,FLP20/FRPR30,randi2023neural,mountoufaris2024line}, as well as learning and memory \cite{melzer2021bombesin}.  In particular, neuropeptide release is typically activity-dependent, and once secreted these molecules act on any nearby neurons expressing relevant receptors. These modulators can alter neuronal dynamics in two distinct ways, either by directly affecting synaptic strength or by altering neuronal properties such as input-output relationships \cite{nadim2014neuromodulation}. 
While synaptic modulation has been widely studied both computationally and experimentally, the effect of neuromodulators on neuronal conductivity has not been assessed. Specifically, how and if neuromodulators affect the capacity of the nervous system for memory retrieval and computability is unknown.

Here, we attempt to address this question by considering a minimal, yet biologically grounded model of neuromodulation. Based on experimental evidence that neuromodulation often acts multiplicatively, effectively gating the neuronal inputs or regulating the integration timescales of individual neurons{\color{\editcolor}~\cite{Marder-Thirumalai02:rev,marder2012neuromodulation, grossman2022neuromodulation,Zeraati-Levina-Macke-Gao26-NatNeurosci}}, we consider a minimal model system that operates as a coupled two-layer network: a primary layer of neuronal activity and an auxiliary layer of neuromodulators that multiplicatively regulate the primary dynamics. 
This concept of multiplicative gating is also canonical in machine learning RNNs. Architectures such as Long Short-Term Memory (LSTM)~\cite{Hochreiter-Schmidhuber97, Gers-Schmidhuber-Cummins99} and Gated Recurrent Unit (GRU)~\cite{Cho--Bengio14,Ravanelli-Brakel-Omologo-Bengio18} utilize gating specifically to control information flow and emergent timescales. 


\par
While {\color{\editcolor}a gated random RNN with otherwise-chaotic dynamics has been shown to robustly self-organize into marginally stable states}~\cite{Krishnamurthy-Can-Schwab22,Can-Krishnamurthy25}, the implications of this self-adaptive gating for \textit{associative memory} remain largely unexplored. 
Using a combination of direct many-body simulations and Dynamical Mean-Field Theory (DMFT), we show that gating fundamentally alters the memory retrieval dynamics. 
Our main finding is that the gated network can bypass the
\deleted{standard sharp transition to the spin-glass phase}%
{\color{\editcolor}sharp loss of retrieval (the ``memory cliff'') at the classical critical capacity $\alpha_\text{c}$}.
{\color{\editcolor}Above $\alpha_\text{c}$, where a cued pattern otherwise survives only as a high-overlap transient~\cite{Amari-Maginu88,Gardner-Derrida-Mottishaw87,clark2025transientdynamicsassociativememory}, multiplicative gating renders this transient retrieval \emph{permanent} when the gates act as fully open or fully closed switches, while graded gating substantially \emph{prolongs} its lifetime, potentially past any physiologically relevant timescales.}
Other instances of autoassociative models with higher-order interactions, known as dense associative memories, have also shown increased storage capacities, algebraically~\cite{Lee86, Gardner87, Abbott-Arian87, Horn-Usher88, Krotov-Hopfield16,Krotov-Hopfield18} or even exponentially large~\cite{Demircigil--Vermet17,Ramsauer--Hochreiter21, Lucibello-Mezard24}\deleted{, but at the expense of narrowed empirical basin sizes~\cite{Lucibello-Mezard24,mimura2025dynamicalpropertiesdenseassociative,clark2025transientdynamicsassociativememory}}. 
{\color{\editcolor} Here, in contrast, the stored pairwise Hebbian couplings remain unchanged, 
as the gate acts on the kinetics, slowing the escape from the ``ghosts'' of stored patterns and, when the gating is strictly on--off, arresting it entirely at a new fixed point set by the input. }
\deleted{We identify the microscopic mechanism: adaptive gating slows down the escape from the ``ghosts'' of stored patterns, eventually stabilizing them into true fixed points.}
Finally, we reveal how gating can lead to a novel memory retrieval phase of continuous multistability, where the system supports a continuous manifold of stable fixed-point attractors.

The remainder of the paper is organized as follows: 
In Sec.\ts\ref{sec:model}, we define the self-adaptively gated associative memory network and discuss its biological and computational motivations, followed by a brief review of the baseline properties of the standard ungated network.
In Sec.\ts\ref{sec:results}, we present the main phenomenological results from direct numerical simulations, including the extended memory retrieval phase. 
In Sec.\ts\ref{sec:dmft}, we develop the DMFT formalism for the gated network to establish that the findings from finite-size simulations persist in the thermodynamic limit, and we analyze the resulting long-time behavior to characterize steady-state overlaps. 
In Sec.\ts\ref{sec:multistability}, we then visualize and characterize the geometry of the emergent continuous multistable attractors in a simplified system storing only two patterns.
Finally, in Sec.\ts\ref{sec:conclusions}, we summarize our findings and discuss their implications for both biological systems and neuromorphic computing. 


\section{MODEL AND BACKGROUND} \label{sec:model}
\begin{figure*}[t]
    \centering
    \includegraphics[width=\textwidth]{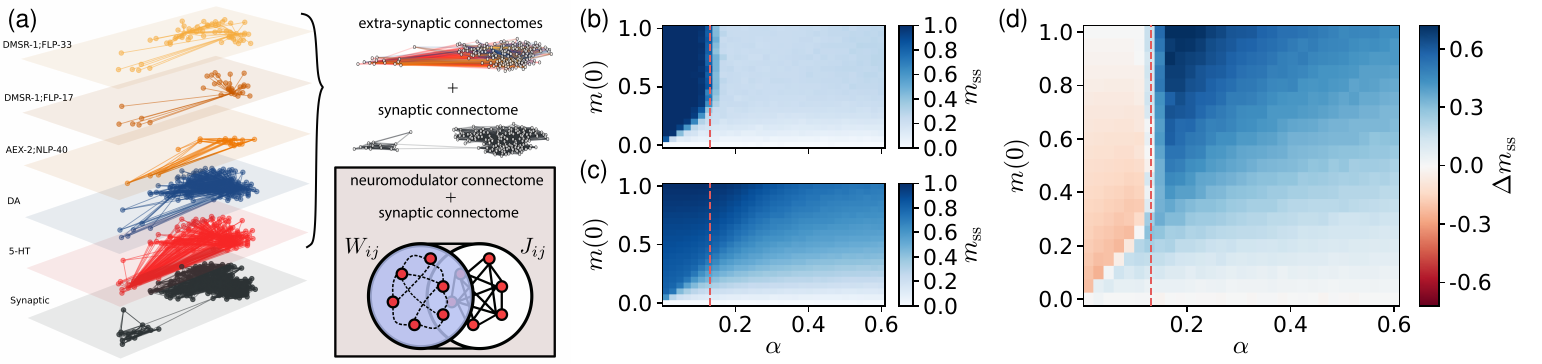}
    \caption{ 
    \label{fig:main}
    {\bf Gating enables memory retrieval beyond the traditional capacity threshold \deleted{without narrowing empirical attractor basins}.} 
    (a) Multilayer neuromodulator networks of \textit{C. elegans} with bioamines serotonin (5-HT, red) and dopamine (DA, blue), and 3 selected neuropeptide-receptor pairs (orange) \cite{ripoll_neuropep_celegans}. These overlay the synaptic connectome (black) \cite{Cook2019Wholeanimal}.
    In the shaded box, we show the schematic of the {\color{\editcolor} model's architecture, where a proxy for the neuromodulatory connectome $W_{ij}$ overlays the synaptic connectivity $J_{ij}$ on the same set of neurons, the two interacting only through neuronal activity}. 
    (b) {Ungated benchmark.}  
    The steady-state overlap $m_\text{ss}$ plotted as a function of memory load $\alpha$ and initial overlap $m(0)$ for the standard Amari--Hopfield network ($\gamma=0$). 
    Color intensity represents the retrieval quality, ranging from failure (white) to successful retrieval (darkblue). 
    The vertical dashed line marks the standard critical capacity $\alpha_\text{c} \simeq 0.13$ of the ungated network. 
    Notably, catastrophic blackout (the spin-glass transition) is observed for $\alpha> 
    \alpha_\text{c}$.
    (c) {Self-adaptively gated network.} 
    Corresponding phase diagram for the same range of parameters in the binary gating limit ($\gamma \to \infty$). 
    In contrast to (b), catastrophic blackout is \emph{not} observed.
    (d) {Performance gain.} 
    The difference in steady-state overlaps, $\Delta m_\mathrm{ss} = m^{\gamma\to\infty}_\mathrm{ss} - m^{\gamma=0}_\mathrm{ss}$, highlighting the region where gating recovers memories that fail in the ungated network. 
    \deleted{This demonstrates that the capacity gain is achieved \emph{without} degrading the size of the attractor basins.}
    Parameters used: $N=1000$, $dt=0.2$, total run time $T=2000$.
    }
\end{figure*}
\subsection{Self-adaptively gated associative memory network}
We study a continuous-state (graded-response) associative memory network of Amari--Hopfield type~\cite{Amari72, Hopfield84} augmented with a multiplicative gating mechanism.
This mechanism is meant to mimic the effect of neuropeptides or neuromodulators on individual neurons~\cite{nadim2014neuromodulation,jekely2024nonsynaptic}. An example of such modulation was found in \textit{C. elegans} where signaling by a neuropeptide changed the voltage dependence of a T-type calcium channel thus impacting conductivity in the neuron~\cite{zang_peptide-gating}. Overall, while the detailed action and chemical and biophysical effects of these molecules are highly varied~\cite{marder2012neuromodulation,beyond_connectome,hevesi2025neuropeptides}, we adopt a minimally simplified model. We focus on the modulation of neuronal conductivity, integration timescales and the neuron input-output function, and ignore neuromodulatory effects on synaptic parameters that have been explored elsewhere \cite{palacios2019neuromodulation}. As the release of neuropeptides requires neuronal activity~\cite{van2012neuropeptide,klose2021temporally}, we couple the strength of neuromodulatory effects to the activity in the neuronal layer. 
Also, following the extensive neuropeptide characterization in model organisms like \emph{C.~elegans}~\cite{ripoll_neuropep_celegans,celegans_neuropep_rev}, we assume a varied collection of neuropeptides that can be emitted at a source neuron and influence the conductivity properties of a target neuron. These minimal assumptions are reflected in our model which consists of two recurrently connected layers (see Fig.\ts\ref{fig:main}(a) for a schematic): the first layer of $N$ neurons with activity (in spiking models: firing rate) $\mathbf{x}(t)$ and the second layer of $N$ auxiliary units $\mathbf{z}(t)$ that mimic neuropeptides or neuromodulators. {\color{\editcolor} Note that this model does not include direct interactions within the neuromodulatory layer $\mathbf{z}(t)$, which would be required to represent peptidergic cascades in which one neuropeptide controls the release of another~\cite{ripoll_neuropep_celegans,celegans_neuropep_rev}.} 

Both state variables are real-valued functions of time  $t\geq0$, that is, $\mathbf{x},\mathbf{z}\,\colon\mathbb{R}_{\geq0}\to\mathbb{R}^N$. 
The neuronal layer is modeled as a fully connected (all-to-all) recurrent network with symmetric Hebbian couplings~\cite{Hebb49:book, Hopfield82}
\begin{equation}\label{eq:hebb}
    J_{ij} = \sum_{\mu=1}^P\xi_i^\mu\xi_j^\mu \ ,
\end{equation}
where each pattern $\bm{\xi}^\mu \in \mathbb{R}^N$ represents a memorized configuration in the neuronal layer, and $P$ is the number of stored patterns.
The memory load $\alpha = P/N$ determines the rank of the coupling matrix $ (J_{ij})_{1\leq{i,j}\leq{N}}$; in particular, for $\alpha < 1$ the matrix is of low-rank (at most $P$)~\cite{Marcenko-Pastur67}, positioning this system within the broader class of low-rank RNNs~\cite{Mastrogiuseppe-Ostojic18}.
Throughout this work, we focus on order-one loads, $\alpha = \mathcal{O}(1)$.
Roman letters ($i,j,\ldots\in\{1,\ldots,N\}=\vcentcolon[N]$) index neurons and neuromodulators, and Greek letters $(\mu,\nu,\ldots\in[P])$ index stored patterns.
\par
The dynamics of the neurons follow the equation:
\begin{equation}\label{eq:gRNN_x}
    \partial_t{x}_i = \sigma(z_i)\bigg[-x_i+\frac{g}{\sqrt{PN}}\sum_{j=1}^NJ_{ij}\phi(x_j)\bigg] \ ,
\end{equation}
where $\phi\,\colon\mathbb{R}\to(-1,1)$ is a bounded nonlinear activation function and $g$ is a scalar {\color{\editcolor} neural} gain.
The prefactor $1/\sqrt{PN}$ in the interaction term ensures that it remains of order one in the thermodynamic limit \cite{clark2025transientdynamicsassociativememory}.
A bounded, monotone nonlinear mapping $\sigma\,\colon\mathbb{R}\to[0,1]$ represents a switch-like multiplicative neuromodulatory gate.
Its value is adaptively controlled by the neuromodulator ${z}_i(t)$, whose time-evolution is recursively driven by the activated neuronal output $\{\phi(x_j)\}_{j=1}^N$ as
\begin{equation}\label{eq:gRNN_z}
    \tau_z\partial_t{z}_i = -z_i + \frac{1}{\sqrt{N}}\sum_{j=1}^NW_{ij}\phi(x_j) \ ,
\end{equation}
where $\tau_z$ is the intrinsic neuromodulator release time-constant, and the neuromodulator-neuron couplings $(W_{ij})_{1\leq{i,j}\leq{N}}$ are modeled as quenched random variables independent of the stored patterns.

\subsubsection{Functional forms and parameterization}
Throughout the paper, we use the following specific choices for the functional forms: 
The neuronal activation and the neuromodulatory gating functions are assumed, respectively, to be $\phi(x)=\tanh(x)$ and $\sigma(z) = (1+\mathrm{e}^{-\gamma z})^{-1}$, where $\gamma\geq0$ sets the steepness of the logistic sigmoid.
We expect that the main results of the paper are robust to quantitative changes for many other common choices of monotonic, integrable functional forms.
Using a nonmonotonic activation function is known to drastically alter associative memory dynamics~\cite{Morita93,kabashima2025dynamicalmeanfieldapproach}, and is not considered here. 

With respect to model parameters, we fix the intrinsic time-constant of the neuromodulating layer and the scalar {\color{\editcolor} neural} gain: $\tau_z=1$ and $g=1.5$. 
The stored patterns are taken to be independent and identically distributed (iid) random variables with zero mean and unit variance.
Unless otherwise stated, we assume binary patterns $\xi_i^\mu\sim\mathsf{Unif}(\{-1,1\})$; 
we have checked numerically that replacing them with Gaussian patterns $\xi_i^\mu\sim\mathcal{N}(0,1)$ leaves the results qualitatively unchanged, consistent with the universality of the spectral properties of large random Wishart matrices~\cite{Tao-Vu12}. 
The neuromodulatory units are initialized with iid standard normal vector, $\mathbf{z}(0)\sim\mathcal{N}(\bm{0},I_N)$.
{\color{\editcolor} As the simplest baseline choice, we take the} inter-layer interactions $(W_{ij})$ {\color{\editcolor} to be} iid random variables, $W_{ij}\sim\mathcal{N}(0,1)$, {\color{\editcolor} with no correlations between matrix elements~\cite{Sommers-Crisanti-Sompolinsky-Stein88}}. {\color{\editcolor} Thus, $W$ is generically asymmetric}, a standard choice in modeling random neural networks~\cite{Sompolinsky-Crisanti-Sommers88, Engelken-Wolf-Abbott23,Krishnamurthy-Can-Schwab22,Can-Krishnamurthy25}. 
\par
Finally, the memory load $\alpha$, gating steepness $\gamma$, and initial neural state $\mathbf{x}(0)$ serve as control parameters for the model defined by Eqs.\ts(\ref{eq:hebb}--\ref{eq:gRNN_z}). 
We note that for these choices, we observe \emph{no} chaotic behavior in the dynamics.
{\color{\editcolor}The roles of $\tau_z$, $\mathbf{z}(0)$, and the rank and sparsity of $W$ are examined in the SI.}

\subsubsection{Theoretical and biophysical motifs}
The value of the gate $\sigma(z_i)$ dynamically and adaptively controls the effective time-constant of the $i^\text{th}$ neuron~\cite{Krishnamurthy-Can-Schwab22, Can-Krishnamurthy25}.
As $\sigma(z_i)$ decreases, the effective integration time of the neuron becomes slower; when $\sigma(z_i)=0$ the state $x_i$ ceases to update and becomes closed.
This specific gating mechanism is analogous to (one of) the canonical gating units originally introduced empirically as a core component of machine learning RNN architectures, such as Long Short-Term Memory (LSTM)~\cite{Hochreiter-Schmidhuber97, Gers-Schmidhuber-Cummins99} and Gated Recurrent Unit (GRU)~\cite{Cho--Bengio14,Ravanelli-Brakel-Omologo-Bengio18}, primarily to avoid vanishing/exploding gradient issues. 
However, Tallec and Ollivier axiomatically proved that the gating-like mechanism necessarily follows from postulating (quasi-) invariance to nonlinear time transformations in the processed data~\cite{Tallec-Ollivier18} (see also~\cite[Ch.5]{bronstein2021geometric}).
Furthermore, in biological neural networks, an equivalent control over a neuron’s effective integration time-constant can arise from various biological mechanisms, such as synaptic shunting inhibition~\cite{Guetig-Sompolinsky09} and negative derivative feedback~\cite{Lim-Goldman13, Can-Krishnamurthy25} in addition to the neuromodulatory regulation we consider here~\cite{Marder-Thirumalai02:rev,marder2012neuromodulation,Alonso-Rue-Marder23}.
{\color{\editcolor} In this light, we expect the gate $\sigma(z_i)$  to be a reasonably grounded biophysical heuristic, positioning our framework as \emph{a minimal model} of {self-adaptively gated associative memory networks}.}

\subsection{Review of the ungated network} \label{sec:review_ungated}
In the limiting case of $\gamma=0$, the gating function becomes constant, $\sigma\equiv 1/2$, at all times regardless of $\mathbf{z}(t)$, reducing the model to the standard associative memory network without gating~\cite{Amari72,Hopfield84,Kuehn-Boes-Hemmen91,clark2025transientdynamicsassociativememory}.  
Before analyzing the gated case, we briefly review the properties of this ungated limit to establish a baseline for comparison.
\par
To quantify the alignment of the neuronal activations at time $t$ with the stored patterns, we introduce the \emph{overlaps} (also called Mattis magnetizations, by analogy with spin glasses~\cite{Amit-Gutfreund-Sompolinsky85a,Amit-Gutfreund-Sompolinsky85b,Amit-Gutfreund-Sompolinsky87}):
\begin{equation}\label{eq:orderparam_def}
    m^\mu(t) = \frac{1}{N}\sum_{i=1}^N \xi_i^\mu \phi(x_i(t)) \quad (\mu\in[P]) \, ,
\end{equation}
which serve as macroscopic order parameters for the retrieval dynamics.
We often focus on the steady-state overlaps $m_\text{ss}^\mu \vcentcolon= \lim_{t\to\infty} m^\mu(t)$ resulting from initial states close to the stored patterns.
A steady-state overlap of order one, $m_\text{ss}^\mu=\mathcal{O}(1)$, indicates successful memory retrieval, whereas $m_\text{ss}^\mu=\mathcal{O}(1/\sqrt{N})$ indicates failure {\color{\editcolor} in the static, equilibrium description~\cite{Amit-Gutfreund-Sompolinsky85a,Amit-Gutfreund-Sompolinsky85b,Amit-Gutfreund-Sompolinsky87}.
Under the cue-conditioned deterministic zero-temperature dynamics considered here, however, failure need not imply an $\mathcal{O}(1/\sqrt{N})$ overlap: a trajectory initialized from a condensed cue can instead terminate at a spurious metastable minimum~\cite{Gardner86} that retains a finite, cue-dependent remanent overlap
with the target pattern~\cite{Gardner-Derrida-Mottishaw87}.}

In the remainder of the paper—except in Sec.\ts\ref{sec:multistability}, where we address the two-pattern case—we focus, without loss of generality, on the retrieval of a single target condensed pattern $\bm{\xi}^1$.
This corresponds to initial conditions where $m^1(0)=\mathcal{O}(1)$ and $m^\mu(0)=\mathcal{O}(1/\sqrt{N})$ for all $\mu\ge2$.
Accordingly, we simply write $m(t)\equiv m^1(t)$, dropping the index, as the overlaps for the remaining patterns satisfy $m^\mu(t)=\mathcal{O}(1/\sqrt{N})$ throughout the dynamics.
\par
In Fig.\ts\ref{fig:main}(b), we show steady-state overlaps computed from direct many-body simulations of Eq.\ts\eqref{eq:gRNN_x} with $\gamma=0$ and $N=1000$. 
The equation was integrated using the Euler method with a time step $dt=0.2$.
The network was initialized as $\mathbf{x}(0)=c_1\bm{\xi}^1+\bm{\eta}$, where $\bm{\eta}\sim\mathcal{N}(\bm{0},I_N)$ and the coefficient $c_1$ is determined numerically to satisfy the target initial overlap $m(0)$. 
Each pixel represents $m(t=2000)$, a time sufficient for the system to relax to a steady state, for a given memory load $\alpha$ and initial overlap $m(0)$, averaged over $100$ realizations of $\{\bm{\xi}^\mu\}_{\mu=1}^P$ and $\mathbf{x}(0)$.
The vertical dashed line marks the \emph{critical capacity}: for $g=1.5$, this occurs at $\alpha_\text{c}\simeq 0.13$~\cite{clark2025transientdynamicsassociativememory}.
In the heatmap, the domain characterized by $m_\text{ss}\simeq1$ (darkblue) defines the memory retrieval phase, where the stored pattern corresponds to a stable fixed point of the collective network dynamics.
However, retrieval fails if either the memory load is too high ($\alpha>\alpha_\text{c}$) or the system is initialized too far from the target pattern ($m(0)< m_\text{c}$), where $m_\text{c}(\alpha)$ is defined by the lower boundary separating the retrieval and failure domains.
Consequently, this plot effectively visualizes the empirical basin of attraction, a concept rigorously analyzed in the context of statistical neurodynamics~\cite{Amari-Maginu88, Okada95}.
\par
According to statistical physics analyses of the static energy landscape based on the replica method~\cite{Amit-Gutfreund-Sompolinsky85a,Amit-Gutfreund-Sompolinsky85b,Amit-Gutfreund-Sompolinsky87,Gardner86}, the region above the critical capacity is a spin-glass phase populated by {\color{\editcolor}the} metastable “spurious” minima, whose configurations coincide with none of the stored patterns{\color{\editcolor} --- though, as noted after Eq.\ts\eqref{eq:orderparam_def}, a minimum reached from a condensed cue can retain a finite remnant overlap with the target}; in this phase the memorized patterns are \emph{no longer} stable attractors.
However, dynamical studies indicate that even above $\alpha_\text{c}$, traces of the stored patterns persist in the energy landscape~\cite{Amari-Maginu88,Morita93,clark2025transientdynamicsassociativememory}, remaining locally stable almost everywhere except in narrow unstable ravines (see, {e.g.}, Fig.\ts4 in~\cite{Morita93}).
Due to this complex residual structure, which is reminiscent of ``ghost'' attractors~\cite{Sussillo-Barak13, Koch--Koseska24,dinc2025ghostmechanismanalyticalmodel} ({i.e.}, regions of slow dynamics emerging amidst the ruins of a destabilized fixed point), even when retrieval ultimately fails, the system trajectory typically approaches the target pattern and transiently lingers in its vicinity with an order-one overlap (as illustrated by the $\gamma=0$ curve in Fig.\ts\ref{fig:transient_retrieval}({\color{\editcolor}a})).
This \emph{transient retrieval} feature~{\color{\editcolor}\cite{Amari-Maginu88,Gardner-Derrida-Mottishaw87,Morita93,clark2025transientdynamicsassociativememory}}, often overlooked in static spin-glass analyses, suggests that associative memory dynamics possesses a typical optimal readout time~\cite{clark2025transientdynamicsassociativememory}, or can even be fundamentally improved by driving the network dynamics out of equilibrium, as exemplified by networks with nonmonotonic activation functions~\cite{Morita93,Yoshizawa-Morita-Amari93,kabashima2025dynamicalmeanfieldapproach}.
\par
In the subsequent sections, we show that a gated network, {\color{\editcolor} in the binary gating limit $\gamma \to \infty$}, governed by Eqs.\ts(\ref{eq:hebb}--\ref{eq:gRNN_z}) can bypass this sharp order--to--spin-glass transition, extending the $m_\text{ss}=\mathcal{O}(1)$ retrieval phase beyond the standard critical capacity threshold $\alpha_\text{c}$. 

\section{MAIN RESULTS} \label{sec:results}
In this section, we present the memory retrieval properties of gated networks obtained from direct numerical simulations of the many-body dynamics governed by Eqs.\ts(\ref{eq:hebb}--\ref{eq:gRNN_z}).
We begin in Sec.\ts\ref{sec:m_ss} by establishing the steady-state retrieval limits, demonstrating how the system circumvents catastrophic failure typical of overloaded networks. 
Next, in Sec.\ts\ref{sec:timescale_by_gate}, we uncover the microscopic origin of this enhancement by analyzing the temporal trajectories, showing how the gate stabilizes otherwise transient dynamics.
These numerical findings are subsequently corroborated in Sec.\ts\ref{sec:dmft} using Dynamical Mean-Field Theory (DMFT) using the bipartite cavity approach~\cite{clark2025transientdynamicsassociativememory}, which establishes that the observations made in finite-size simulations persist in the thermodynamic limit and in the long-time limit.
Finally, in Sec.\ts\ref{sec:multistability}, we investigate the emergent continuous multistability that naturally arises from the self-adaptive gating mechanism, visualizing the attractor landscape in a simplified setting.
\par

\subsection{Extended memory retrieval phase} \label{sec:m_ss}
In the previous section, we reviewed that in the absence of gating ($\gamma=0$), the associative memory network exhibits a sharp phase transition between the memory-retrieval phase and the spin-glass phase.
Using the same parameters, we examine steady-state overlaps of the gated networks in the binary gating limit ($\gamma\to\infty$), as shown in Fig.\ts\ref{fig:main}(c).
In stark contrast to the ungated case, the sharp, discontinuous transition where memory retrieval suddenly fails is notably absent. 
Instead, the gated network maintains an order-one retrieval overlap well beyond the critical capacity of the ungated network, $\alpha_c \simeq 0.13$.
\par
However, this expansion of the retrieval domain comes with a trade-off regarding retrieval fidelity.
To quantify this, we plot the difference in steady-state overlaps between gated and ungated networks, 
\begin{equation}
    \Delta m_\text{ss} = m_\text{ss}^{\gamma\to\infty} - m_\text{ss}^{\gamma=0} \ ,
\end{equation}
in Fig.\ts\ref{fig:main}(d). 
Below critical storage capacity ($\alpha < \alpha_\text{c}$), the difference $\Delta m_\text{ss}$ is dominated by negative (red) to near-zero (white) values. 
This indicates that while the gated network still retrieves
patterns, the gating mechanism slightly degrades the fidelity (magnitude) of the final overlap compared to the ungated baseline.
Conversely, in the memory overload regime ($\alpha>\alpha_\text{c}$),  strongly positive (blue)  $\Delta m_\text{ss}$ dominate. 
Here, gating enables successful memory retrieval in regions where the ungated network fails completely ({i.e.}, where $m_\mathrm{ss}^{\gamma=0}\approx0$). 
\par
{\color{\editcolor}The dependence of the empirical retrieval boundary $m_\text{c}$ on the cue construction, the required retrieval fidelity, and the system size is characterized in SI Sec.\ts\ref{sec:cue-construction}; the extension of retrieval beyond $\alpha_\text{c}$ is independent of these choices.}
\deleted{Fig.\ts\ref{fig:main} also shows that the critical retrieval boundary $m_\text{c}(\alpha)$ remains largely unchanged, indicating that this extended retrieval does not come at the expense of shrinking the empirical attractor basins.}
\deleted{This property stands in contrast to dense associative memory models~\cite{Lee86, Gardner87, Abbott-Arian87, Horn-Usher88, Krotov-Hopfield16,Krotov-Hopfield18}, which achieve huge capacity gains scaling algebraically in $N$ but typically suffer from the narrowing of empirical basins as the interaction order $n$ increases, manifested by an upward shift of the critical retrieval boundary $m_\text{c}(\alpha)$~\cite{mimura2025dynamicalpropertiesdenseassociative,clark2025transientdynamicsassociativememory}.}

\subsection{Dynamics of pattern retrieval in the gated networks} \label{sec:timescale_by_gate}
\begin{figure}[!t]
    \centering
    \includegraphics[width=0.45\textwidth]{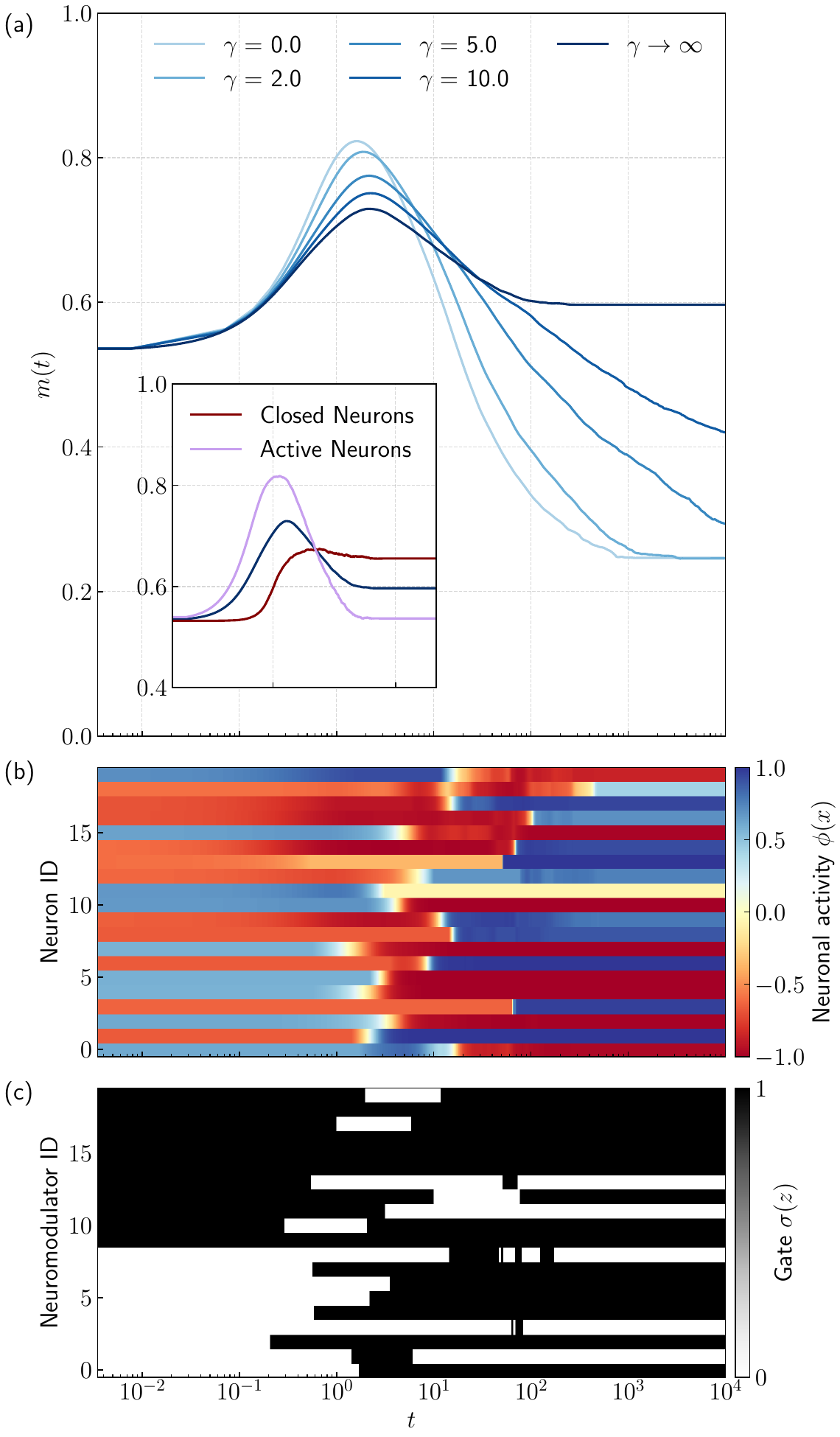}
    \caption{\label{fig:transient_retrieval}
    \textbf{Gating stabilizes retrieval of patterns in regimes beyond critical capacity}
    (a) The temporal evolution of the overlap $m(t)$ in the memory overload regime ($\alpha=0.4 > \alpha_\text{c}$) for various gating steepness parameters $\gamma$. 
    The case $\gamma=0$ corresponds to the standard ungated network, which exhibits a transient high-overlap state that eventually decays.
    As $\gamma$ increases, the lifetime of this transient extends, eventually leading to a stable retrieval state.
    Each curve is averaged over 30 independent simulation runs with different realizations of the patterns $\{\bm{\xi}^\mu\}_{\mu=1}^P$, couplings $W_{ij}$, and initial conditions $\mathbf{z}(0)$ and $\mathbf{x}(0)$ (fixed at $m(0)\simeq0.55$).
    (Inset) Decomposition of the overlap for the binary gated limit ($\gamma\to\infty$) into the closed ($m_{\mathcal{F}_t}(t)$) and active ($ m_{\mathcal{A}_t}(t)$) sub-population overlaps.
    (Bottom) Representative dynamical traces of neuronal activity $\phi(x_i)$ (b) and gating values $\sigma(z_i)$ (c) for the binary gated case, illustrating the dynamical exchange between frozen and active populations before reaching the steady state.
    Parameters used: $N=1000$ and $dt=0.2$.
    }
\end{figure}
In this section, we analyze the dynamics to gain insight into the mechanism underlying the extended memory retrieval observed in the phase diagrams. 
We focus on the temporal evolution of the overlap $m(t)$ in the overload regime ($\alpha=0.4 > \alpha_\text{c}$), where the standard ungated network fails to retrieve.
Fig.\ts\ref{fig:transient_retrieval}(a) displays the trajectories of the overlap starting from an initial condition relatively close to the stored pattern ($m(0) \simeq 0.55$).
\par
For the ungated benchmark ($\gamma=0$), the dynamics exhibit the characteristic transient retrieval behavior discussed in Sec.\ts\ref{sec:review_ungated}. 
The system is initially attracted toward the stored pattern, maintaining a high overlap for a finite duration by lingering in a region of slow dynamics~\cite{clark2025transientdynamicsassociativememory}. 
However, because the pattern is not a stable fixed point at this high load, the system eventually escapes this vicinity and relaxes to a spin-glass state (a spurious attractor with small overlap).
\par
Introducing the self-adaptive gating ($\gamma > 0$) fundamentally alters this behavior.
As the gating steepness $\gamma$ increases, the duration of the high-overlap plateau extends significantly. The gating mechanism, particularly in the limit of sharp gating, seemingly stabilizes what was a transient state in the ungated dynamics into a true fixed point. Note that while the results presented in this section are numerical, in a subsequent section, using an asymptotic variant of dynamical mean-field theory, we show that the system does indeed find a true fixed point. 

To further understand the dynamics, we consider the binary gate limit and partition the neuronal population into two distinct states where the gate is closed ($\sigma(z_i(t))=0$), and the population where the gate is open ($\sigma(z_i(t))=1$)~\cite{Can-Krishnamurthy25}. 
Denoting the sets of indices for these populations as $\mathcal{F}_t$ and $\mathcal{A}_t$ respectively, we define the closed and active sub-population overlaps as 
\begin{equation}
    m_{\mathcal{F}_t}(t) = \frac{1}{|\mathcal{F}_t|}\sum_{i\in\mathcal{F}_t}\xi_i^1\phi(x_i) \ , \quad 
    m_{\mathcal{A}_t}(t) = \frac{1}{|\mathcal{A}_t|}\sum_{i\in\mathcal{A}_t}\xi_i^1\phi(x_i) \ ,
\end{equation}
where $|\mathcal{F}_t|$ and $|\mathcal{A}_t|$ denote the number of neurons in each set at time $t$ (with $|\mathcal{F}_t|+|\mathcal{A}_t|=N$).
The total overlap $m(t)$ is thus given by the weighted sum $m(t)=\rho_{\mathcal{F}_t}(t)m_{\mathcal{F}_t}(t)+\rho_{\mathcal{A}_t}(t)m_{\mathcal{A}_t}(t)$, where $\rho_{\mathcal{F}_t}(t)=|\mathcal{F}_t|/N$ and $\rho_{\mathcal{A}_t}(t)=|\mathcal{A}_t|/N$ are the respective population fractions.
\par
The inset in Fig.\ts\ref{fig:transient_retrieval}(a) displays the temporal evolution of these sub-overlaps.
Crucially, while the total overlap $m(t)$ stabilizes, the membership of these populations is not static.
As shown in Fig.\ts\ref{fig:transient_retrieval}(b,c), the two populations dynamically and adaptively exchange
members over time.
This indicates that stabilization of the macroscopic order parameter is mediated by switching between the closed and active neuron subpopulations during relaxation; 
once the dynamics has converged, both the overlap and the subpopulations become stationary. 
{\color{\editcolor}This relaxation, in which a dynamically selected closed (or frozen) subset statically stabilizes the remaining active neurons, is an associative memory realization of the frozen-stabilization mechanism of Can and Krishnamurthy~\cite{Can-Krishnamurthy25}. Frozen-stabilization carries a characteristic signature in the instantaneous Jacobian spectrum, where the frozen coordinates contribute extensive exact zero modes while the active subset becomes stable, and our gated network reproduces this signature during retrieval (SI Sec.\ts\ref{sec:jacobian-spectra}). Moreover, because the mechanism requires only that some subset of neurons be frozen, which the binary gate supplies for a broad range of couplings, retrieval beyond the critical capacity is largely robust to the rank and sparsity of $W$; at large but finite $\gamma$, these properties instead control the lifetime of transient retrieval. We discuss both behaviors in SI Sec.\ts\ref{sec:modulatory-robustness}.}

\section{Thermodynamic limit and time-asymptotic behavior} \label{sec:dmft}
The observations from the finite-size many-body simulations in the previous section raise the following question:
Does the extended memory retrieval induced by gating persist in the thermodynamic limit and the long-time asymptotic regime?
To address this question, we analyze the system using a Dynamical Mean-Field Theory (DMFT) which enables us to obtain analytical expressions based on the statistical properties of the gated networks.
By taking the thermodynamic limit ($N,P\to\infty$ while keeping the ratio $\alpha=P/N$ as order-one $\mathcal{O}(1)$), DMFT effectively maps the high-dimensional many-body dynamics onto a set of low-dimensional, self-consistent stochastic equations describing a representative single-site process for the variables $x(t), z(t)$, and the order parameter $m(t)$.

\subsection{Self-consistent DMFT equations}
Extending the bipartite dynamical cavity method developed for ungated networks in Ref.\ts\cite{clark2025transientdynamicsassociativememory} (see {\color{\editcolor}SI Sec.\ts\ref{sec:si_dmft}} for details), we obtain the effective equations of motion:
\begin{equation}
    \partial_t x = \sigma(z)\bigg[ -x + \frac{g}{\sqrt{\alpha}}\big( \xi m + \eta_x  + {\alpha}h_\mathrm{ret}(t) \big) \bigg] \ ,
    \label{eq:dxdt_dmft}
\end{equation}
where the retarded self-interaction term is given by $h_\mathrm{ret}(t) = \int\!du \, G(t,u)\phi(u)$ with shorthand $\phi(t)=\phi(x(t))$, and 
\begin{equation}
    \partial_t z = -z + \eta_z \ .
\end{equation}
Here, $m(t)=\langle{\xi\phi(t)}\rangle$ is the overlap of the condensed pattern, $\eta_x(t)$ and $\eta_z(t)$ are temporally colored Gaussian noises with zero mean and variances given by
\begin{align}
    \langle{\eta_x(t)\eta_x(t')}\rangle &= \alpha \int\!{du}\!\int\!{du'}\,G(t,u)C_\phi(u,u')G(t',u') \\
    \langle{\eta_z(t)\eta_z(t')}\rangle &= C_\phi(t,t') \ .
\end{align}
Finally, the self-consistency loops are closed by the two-time correlation function
\begin{equation}
    C_\phi(t,t') = \langle{\phi(t)\phi(t')}\rangle 
\end{equation}
and the (retarded) response functions
\begin{equation}
    R(t,t') = \frac{\delta\langle{\phi(t)}\rangle}{\delta h_x(t')} \quad\text{and}\quad G(t,t') \ ,
    \label{eq:G_dmft}
\end{equation}
which obey causality ({i.e.}, $R(t,t')=G(t,t')=0$ for $t<t'$), 
where the latter is defined via the equation
\begin{equation}
    G(t,t') = \delta(t-t') + \int\!{du}\,R(t,u)G(u,t') \ . \label{eq:dyson}
\end{equation}
Here $h_x(t)$ represents the auxiliary external perturbation field, which is introduced to define the response function and is eventually taken to zero.
The delta function in Eq.\ts\eqref{eq:dyson} represents a temporally local perturbation.
The DMFT averages $\langle\dots\rangle$ are taken over the distributions of the effective Gaussian noises $\{\eta_x, \eta_z\}$, the local pattern variable $\xi$, and initial configurations $\{{x}(0), {z}(0)\}$. 
Note that the statistics of these Gaussian noises implicitly capture the ensemble averages over the quenched disorder, including the neuromodulatory couplings $(W_{ij})$ and the non-condensed patterns. 
\par
This closed set of equations (\ref{eq:dxdt_dmft}--\ref{eq:dyson}) defines the non-Markovian process governing the gated associative memory network in the thermodynamic limit. 
Because the order parameters appear within the dynamical expressions, these equations must be solved self-consistently (typically via numerical iteration~\cite{Eissfeller-Opper92,Roy-Biroli-Bunin-Cammarota19}) until the trajectories converge within a specified tolerance.

\begin{figure*}[t]
    \centering
    \includegraphics[scale=0.7]{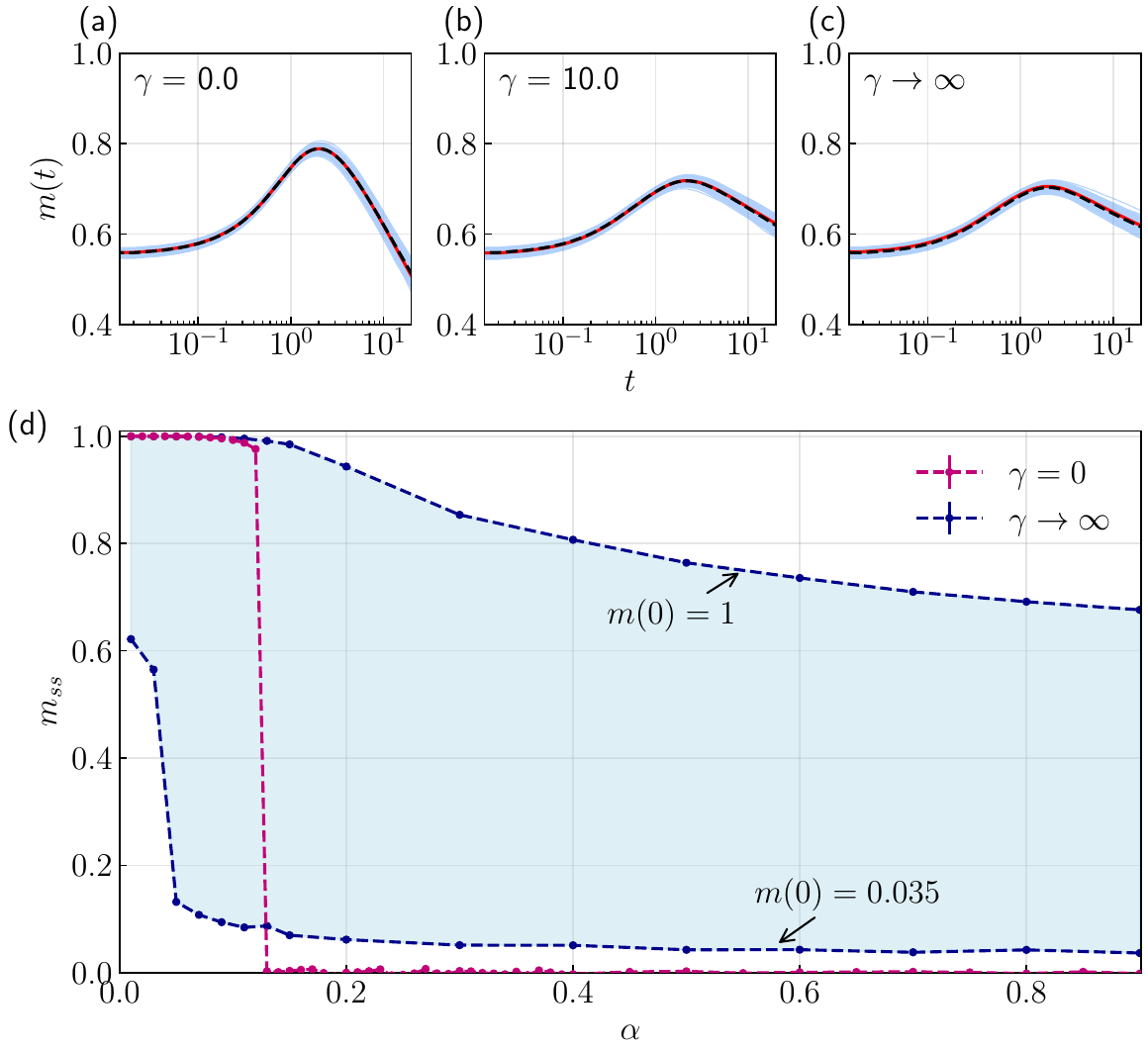}
    \caption{\label{fig:dmft-mft}
    \textbf{DMFT dynamics and their time-asymptotic solutions}.
     (Top row) Overlap dynamics $m(t)$ of ungated and gated networks for memory load above critical capacity $\alpha = 0.4$ and initial overlap $m(0) \simeq 0.65$ using many-body simulations and DMFT: (a) ungated $\gamma=0$, (b) intermediate gating $\gamma=10.0$, and (c) binary gating $\gamma \to \infty$. Light blue curves correspond to individual many-body simulations while the dashed black curves are their average. Solid red curves show the DMFT predictions. (d) Time-asymptotic DMFT solutions for overlaps in ungated (red dots and dashed lines) and gated (blue dots and dashed lines) networks. The band of steady state overlap values for the gated network is bound{\color{\editcolor}ed} by the \deleted{the} maximum and minimum approximate steady state overlap values of the closed neurons $m_{\mathcal{F}_t}^{\text{ss}}$. {\color{\editcolor} The maximum and minimum values of $m_{\mathcal{F}_t}^{\text{ss}}$ were obtained from DMFT calculations integrated to $t=50$ at $m(0)= 1$ and $m(0) = 0.035$, respectively, for the range of memory loads $\alpha$.} DMFT curves in the top row were calculated using the numerical parameters: $M=10000, \alpha=0.4, dt=0.02$. In (d), dashed lines were added to guide the eye.
    }
\end{figure*}

\subsection{Results and Numerical solution of DMFT}
Our DMFT formulation for the gated network agrees well with numerical simulations across a broad range of sigmoid slopes, as shown in Fig.\ts\ref{fig:dmft-mft}(a)--(c).
We observe excellent agreement between the DMFT solutions and the mean trajectories obtained from many-body simulations far above the traditional critical capacity ($\alpha=0.4\gg\alpha_\text{c}$), across a range of sigmoid slopes, despite the non-Gaussian nature expected from the multiplicative gate~\cite{Krishnamurthy-Can-Schwab22,Can-Krishnamurthy25} and crosstalk noise~\cite{nishimori1993retrieval,ozeki1994noise}.
\par
Having validated the DMFT equations in the dynamical regime {\color{\editcolor} (see SI Sec.\ts\ref{sec:si_dmft_nongauss} for additional analysis)}, we now examine their asymptotic properties in the long-time limit. Due to the $\mathcal{O}(T^2)$ time complexity for each iteration of the DMFT curves~\cite{clark2025transientdynamicsassociativememory}, obtaining time-asymptotic curves using the usual DMFT iteration scheme is computationally prohibitive. 
To this end, we derive time-asymptotic DMFT equations by assuming that the dynamics reach stable fixed points $\partial_tx=0, \partial_tz=0$ at $t\to \infty$. 
This yields a set of algebraic equations that are solved self-consistently to obtain time-independent fixed-point values of the gated network's order parameters; see {\color{\editcolor}SI Sec.\ts\ref{sec:asymptotic_dmft}} for more details. 

Fig.\ts\ref{fig:dmft-mft}(d) displays the resulting band of steady-state overlap values $m_{\text{ss}}$, calculated from the time-asymptotic DMFT expressions. 
Each point in the band is parameterized by an imposed overlap value specific to neurons with closed gates, specifically, their approximate steady state overlap values $m_{\mathcal{F}_t}^{\text{ss}}$. Maximum and minimum values of $m_{\mathcal{F}_t}^{\text{ss}}$ will form the bounds of $m_{\text{ss}}$. We therefore constructed the bounds of Fig.\ts\ref{fig:dmft-mft}(d) by obtaining $m_{\mathcal{F}_t}^{\text{ss}}$ from DMFT calculations at $t = 50$ for the extreme cases of maximum initial overlap $m(0) = 1$, and  minimal overlap $m(0) = 0.035$ for different values of memory load $\alpha$.
This analysis suggests that the closed-gate subpopulation neurons prevent active neurons from fully escaping the transient attractors in the landscape near the stored memory.  

\section{EMERGENT CONTINUOUS MULTISTABLE ATTRACTORS AND STABILIZATION OF PATTERNS BEYOND CRITICAL CAPACITY} \label{sec:multistability}
\begin{figure*}
    \centering
    \subfloat[Ungated ($\gamma=0$)\label{fig:phase_flow_ungated}]{
        \includegraphics[width=0.4\textwidth]{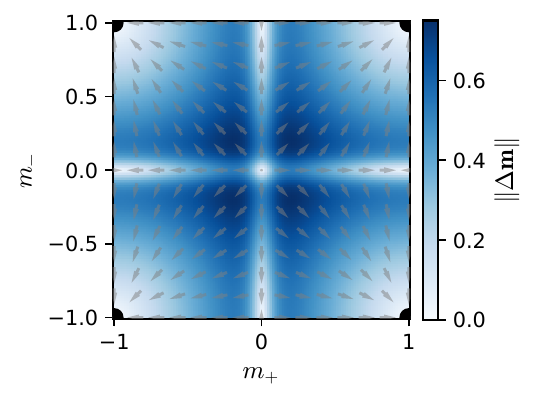}
    }
    \qquad 
    \subfloat[Gated ($\gamma\to\infty$)\label{fig:phase_flow_gated}]{
        \includegraphics[width=0.4\textwidth]{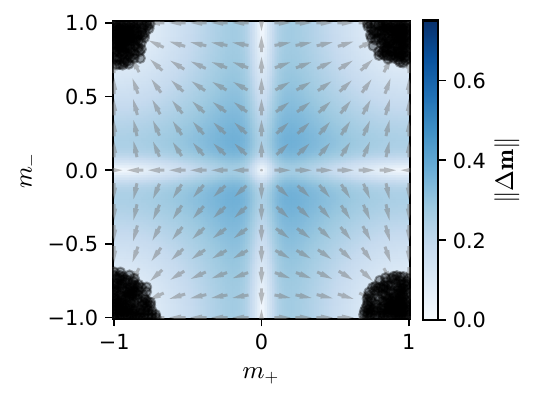}
    }
    
    \caption{\label{fig:phase_flow}
    \textbf{Gating creates continuous manifolds of fixed points.}
    Flow diagrams in the overlap phase space $\tilde{\mathbf{m}}=(m_+,m_-)$, where $m_\pm=m_1\pm m_2$, for a network storing $P=2$ orthogonal patterns.
    Arrows represent the flow direction of the vector field, and dark circles indicate stable fixed points reached from various initial conditions.
    The background heatmap encodes the displacement magnitude $\|{\Delta\mathbf{m}}\|=\|\mathbf{m}(T)-\mathbf{m}(0)\|$ with $\mathbf{m}=(m_{1},m_{2})$, serving as a proxy for the attractor landscape.
    (a) In the {ungated network}, the system flows into one of four discrete fixed points (corresponding to the retrieval of $\bm{\xi}^1, \bm{\xi}^2$ or their inverses), determined solely by the initial basin of attraction.
    (b) In the {gated network}, the system reaches a continuum of distinct fixed points depending on the specific initial overlap, creating a ``cloud'' of stable fixed points.
    For both panels, the realization of patterns $\{\bm{\xi}^1,\bm{\xi}^2\}$, couplings $(W_{ij})$, and  initialization $\mathbf{z}(0)$ are identical and fixed; only $\mathbf{x}(0)$ varies.
    Parameters used: $N=1000$, $dt=0.25$, and $T=2000$.
    }
\end{figure*}
The observations from direct numerical simulations in Sec.\ts\ref{sec:results} and the subsequent DMFT calculations have established that self-adaptive gating fundamentally alters memory retrieval. We can summarize the phenomenology that leads to improved recovery as follows. As in the ungated case, even above critical capacity $\alpha_c$, the system initially gets drawn to the memory state. In the ungated case, this transient behavior is quickly lost in the regime above critical capacity. In the gated case however, this transient increase in the overlap can be leveraged. A fraction of the neurons with improved overlap with the memory pattern can switch from active to close{\color{\editcolor}d} states in this transient regime. These neurons can provide a guiding field for the other active neurons and ensure that there are no catastrophic failures in retrieval. 
In the language of dynamical landscapes, this mechanism can potentially be understood as the stabilization of ``ghost'' attractors~\cite{Sussillo-Barak13,dinc2025ghostmechanismanalyticalmodel}---regions of slow dynamics that persist near the stored patterns even after the true local minima have been destabilized by memory overload.
This gating-induced stabilization can occur at any points in the slow regions of the original landscape; the stabilization condition can be satisfied at various points along the slow-transient trajectory.
This leads to the hypothesis that a discrete fixed point of the standard ungated network can effectively be replaced by a continuously varied multiple stable fixed points.
{\color{\editcolor}Consequently, the gated network is expected to support a continuum of final overlaps (\emph{multistability}), with the steady-state overlap depending continuously on the initial overlap, $m(0) \mapsto m_\mathrm{ss}$, rather than collapsing onto a stored pattern as in the ungated case. The band of time-asymptotic DMFT solutions in Fig.~\ref{fig:dmft-mft}(d) shows that this continuum survives in the thermodynamic limit. Overall, the DMFT treatment further presumes that $m_\mathrm{ss}$ is a deterministic function rather than a random variable; we verify this assumption directly using many-body simulations in SI Sec.\ts\ref{sec:finite-size}.}
\par
Evidence of this emergent multistability is indeed already apparent in the phase diagrams of Fig.\ts\ref{fig:main}: while the ungated network displays essentially no variation in the steady-state overlap $m_\text{ss}$ for a fixed load $\alpha$, the gated network exhibits graded changes in $m_\text{ss}$, indicating a strong dependence on the initial condition $m(0)$.
Crucially, this is substantiated by the continuous band of $m_\text{ss}$ observed in the asymptotic DMFT analysis (Fig.\ts\ref{fig:dmft-mft}(d)), which implies that the gated network theoretically supports a solution space forming a continuous manifold of stable fixed points.
\par
To obtain further intuition and to verify this hypothesis of multistability, we now visualize the attractor geometry by analyzing the phase flow in a simplified system with only two stored patterns ($P=2$) in the low-load limit.
We set $N=1000$, ensuring the memory load is far below the critical capacity, and construct two orthogonal patterns: $\bm{\xi}^1$ is sampled uniformly from $\{-1, 1\}^N$, and $\bm{\xi}^2$ is constructed by flipping the sign of exactly $N/2$ randomly chosen components of $\bm{\xi}^1$.
This construction guarantees that the two patterns are orthogonal to each other, satisfying $\bm{\xi}^{1\top}\bm{\xi}^2=0$.
In this setup, we track trajectories in the two-dimensional overlap phase space parameterized by $\tilde{\mathbf{m}}=(m_+,m_-)$ with $m_\pm=m_1\pm m_2$.
\par
To confirm that the observed multistability is intrinsic to the gating dynamics and not an artifact of varying quenched randomness, we fix the stored patterns (defining $(J_{ij})$), the inter-layer interactions $(W_{ij})$, and the neuromodulatory initialization $\mathbf{z}(0)$. 
We then vary only the initial neural state $\mathbf{x}(0)$ (and consequently the initial overlap $\mathbf{m}(0)$).
\par
In the ungated network (Fig.\ts\ref{fig:phase_flow}(a)), the overlaps invariably converge to one of four discrete stable fixed points: $\mathbf{m}_\text{ss} \simeq (\pm 1, 0)$ or $(0, \pm 1)$ (equivalently $\tilde{\mathbf{m}}_\text{ss}\simeq(1,\pm1)$ or $(-1,\pm1)$), consistent with the $\mathbb{Z}_2$ equivariance of the dynamical equation~\eqref{eq:gRNN_x} (reflecting the symmetry between retrieving $\bm{\xi}^\mu$ and $-\bm{\xi}^\mu$).
Which of these fixed points the system flows into is determined solely by the attractor basin in which the initial overlap $\mathbf{m}(0)$ resides; for example, trajectories initiated at distinct points, such as $\mathbf{m}(0)=(0.4,0)$ (equivalently $\tilde{\mathbf{m}}(0)=(0.4,0.4)$) and $\mathbf{m}(0)=(0.6,0.3)$ (equivalently $\tilde{\mathbf{m}}(0)=(0.9,0.3)$), both converge to the same fixed point $\mathbf{m}_\text{ss}=(1,0)$.
In stark contrast, the flow in the gated network (Fig.\ts\ref{fig:phase_flow}(b)) reaches a continuous range of fixed points depending on $\mathbf{m}(0)$, forming a ``cloud'' of multistable attractors.
Consequently, the two initial conditions $\mathbf{m}(0)=(0.4,0)$ and $\mathbf{m}(0)=(0.6,0.3)$ settle into distinct fixed points.
While the specific geometry of this attractor manifold reflects the particular realization of the quenched disorder ($J, W, \mathbf{z}(0)$), the existence of continuous multistability is a robust feature of the gating mechanism.
In Fig.\ts\ref{fig:phase_flow}, the background heatmap encodes the magnitude of the displacement, $\|{\Delta\mathbf{m}}\|=\|\mathbf{m}(T)-\mathbf{m}(0)\|$. 
Thus, even without a presupposed global potential (or Lyapunov function) for the gated dynamics, this displacement metric serves as an effective proxy for the attractor landscape.
This observation suggests that in the gated networks, the steady-state overlaps can be continuously controlled by choosing a different initial neural condition.
\par
Lastly, we comment on a conceptual parallel with previous studies on \emph{three-body interactions}, where emergent multistability was observed in coupled phase oscillators~\cite{Tanaka-Aoyagi11} and corresponding associative memory networks~\cite{Ota-Tanaka-Aoyagi13}.
In those frameworks, the continuous manifold of attractors arises from specific static higher-order couplings.
In contrast, our results demonstrate that similar continuous multistability can emerge dynamically from multiplicative gating applied to standard pairwise interactions.
Nevertheless, we anticipate that the gating mechanism inherently induces a certain type of effective higher-order interactions.
This can be seen by considering the adiabatic limit $\tau_z \to 0$~\cite{krotov2021hierarchicalassociativememory} for a network with finite $\gamma$, where the neuromodulatory layer reacts instantaneously to the neuronal layer: $z_i(t) \simeq \frac{1}{\sqrt{N}}\sum_j W_{ij}\phi(x_j)$.
Substituting this into the Taylor expansion of the gating function $\sigma(z_i)$ around zero, the effective dynamics of $x_i$ becomes
\begin{align}
    \partial_t x_i &\simeq \bigg( \frac{1}{2} + \frac{\gamma}{4} z_i + \mathcal{O}(z_i^3) \bigg) \bigg[ -x_i + \frac{g}{\sqrt{PN}}\sum_k J_{ik} \phi(x_k) \bigg] \nonumber \\
    &\sim \cdots + \frac{\gamma g}{4\sqrt{P}N} \sum_{j,k} W_{ij} J_{ik} \phi(x_j) \phi(x_k) + \cdots \ . \nonumber
\end{align}
The cross-term $W_{ij}J_{ik}\phi(x_j)\phi(x_k)$ effectively acts as a three-body interaction involving neurons $i, j,$ and $k$.
While this expansion formally diverges in the binary limit ($\gamma\to\infty$), it qualitatively illustrates how multiplicative gating generates a hierarchy of higher-order correlations.
Elucidating the formal relationship between these dynamically generated interactions and explicit higher-order couplings remains an intriguing direction for future work.
\section{CONCLUSION AND OUTLOOK} \label{sec:conclusions}
In this work, we demonstrated that a minimal model of self-adaptive gating fundamentally enhances the classical associative memory properties.
Specifically, we uncovered three key insights regarding the impact of gating on memory retrieval properties:
First, the gated network circumvents the catastrophic spin-glass transition, {\color{\editcolor}maintaining order-one retrieval overlap} well beyond the standard critical capacity. {\color{\editcolor}This extension carries two costs, quantified in SI Sec.\ts\ref{sec:cue-construction}: retrieval fidelity is reduced relative to the ungated network below critical capacity, and the empirical basin is enlarged only below a threshold overlap, being smaller than the ungated basin at higher thresholds.}
Second, we identified the dynamical origin of this enhancement as the gating-induced stabilization of transients, where the adaptive freezing of neuronal integration times stabilizes the slow regions of dynamics near the stored patterns converting them into stable fixed points {\color{\editcolor}in the binary gating limit $\gamma \to \infty$}. 
Third, the stabilization of transients result{\color{\editcolor}s} in the emergence of continuous multistability, transforming the attractor landscape from a discrete fixed point into a continuous manifold of stable fixed points.
All these findings from direct many-body numerical simulations were corroborated by DMFT, establishing that these phenomenologies are genuine features of gating in the thermodynamic limit and in the long-time limit, rather than artifacts of finite system size or long-lived transient effects.
{\color{\editcolor}The binary limit is an idealization: at finite $\gamma$ the gate never closes completely, so these states are long-lived rather than permanent, with Fig.~\ref{fig:transient_retrieval}(a) showing the lifetime of the high-overlap plateau growing steadily with $\gamma$ toward the binary limit. We regard this finite-$\gamma$ regime as the more biologically apt one, since behavioral states in animals persist over finite timescales. Furthermore, when neuromodulation is slower than neural dynamics $\tau_z > \tau_x$, extended memory retrieval requires an initial or pre-{\color{\editcolor}existing} heterogeneous gate configuration, see SI Sec.\ts\ref{sec:tauz-initialization} for more details.} 
\par
Having established and theoretically verified the effects of {\color{\editcolor}a simplified model of} neuromodulation on memory retrieval, a natural question arises, {\it how applicable are the findings of this minimal model for biological systems?} Across all species investigated, the neuropeptide repertoire is vast, with multiple neuron classes releasing multiple neuropeptides. This sets up an impressive layered complexity, potentially adding substantial computational capacity to small nervous systems such as in nematodes. Even mammalian brains show remarkable diversity of neuropeptide expression, whose local release and uptake affect local neuronal populations. Experimentally, it is hard to assess the impact of these modulators in complex brains, as spatial control, or the ability to image the modulators directly are very limited. Computational work suggests that further efforts in this direction are worthwhile - these modulators may prove to be essential for capturing key aspects of brain function. Indeed, dopamine (a neuromodulator) has been found to lead to increased working memory in the hippocampus by affecting individual neurons in the region~\cite{enhanced_work_mem_dope}. 
Neuropeptides have been shown to prolong and stabilize  behavioral states by signaling specific neurons in {\it C. elegans}~\cite{flavell2013serotonin, bonnard2022automatically}. Furthermore, neuropeptides were reported as necessary for line attractor neuronal activity dynamics in mice behavior~\cite{mountoufaris2024line}. 
\par
While each of the previous examples highlight{\color{\editcolor}s} features potentially related to our findings, nonetheless, our gated RNN should be improved for modeling biological systems. 
Specifically, the neuromodulator-neuron couplings $W_{ij}$ are currently quenched random variables, but based on the neuropeptid{\color{\editcolor}e}\deleted{u}rgic connectome of {\it C. elegans}~\cite{ripoll_neuropep_celegans}, {\color{\editcolor} they should contain structure and low rank. Specific species of neuropeptides would connect specific neuron populations, and interactions between the species are biologically possible. One path towards estimating these effects would be to learn $W_{ij}$ couplings from measurements of neuronal activity.} In particular, {\it C. elegans} would be {\color{\editcolor}a} good candidate for this due to its small set of neurons and its extensive publicly available data covering the activity of the majority of its neurons \cite{kato2015global, hallinen21,atanas2023brain, dag2023dissecting}. 
Additionally, the model could be adapted for spiking networks as it currently uses a continuous state akin to analog neurons or as a proxy for firing rates.   
\par

{\color{\editcolor} Furthermore, the emergence of continuous multistable attractors through gating means that the retrieved state is set continuously by the initial overlap, rather than being quantized to a stored pattern: a feature that can be exploited for input-driven computation with persistent activity~\cite{Seung96,Romo-Brody99,Remington-Jazayeri18,khona2022attractor}. In fact, Costacurta et al.~\cite{costacurta2024structured} found that a variant of the low-rank gated RNN trained on temporal tasks extrapolated beyond their trained parameter range (a form of generalization distinct from performance on a withheld test set)~\cite{beiran-ostojic_mwg} and that the gating layer was essential for correctly classifying inputs outside that range. The utility of gating may extend from biological systems to machine learning more broadly: for example, gated attention has recently been shown to improve expressivity and trainability of large language models~\cite{qiu2025gated}. An interesting direction would be to study how gating affects generalization in generative diffusion models and other forms of neuromorphic computing.}

\par
{\it Note added during manuscript preparation:} During the final preparation of this manuscript, a relevant preprint appeared~\cite{delgaudio2025shorttermplasticityrecallsforgotten} that also concerns memory retrieval above the traditional critical capacity.
While both studies employ two-layer network architectures, the mathematical formulations and physical interpretations differ fundamentally.
In our model, the self-adaptive gating mechanism acts \emph{multiplicatively} on the dynamics (modulating the effective time-constants), whereas the model in Ref.~\cite{delgaudio2025shorttermplasticityrecallsforgotten} employs an \emph{additive} modulation of the synaptic interactions.
Consequently, these approaches map to distinct biophysical motifs: our framework models neuromodulation (specifically, neuropeptidergic regulation), whereas their framework models short-term synaptic plasticity.
Furthermore, the underlying mechanisms are of a distinct nature; while Ref.~\cite{delgaudio2025shorttermplasticityrecallsforgotten} proposes a ``trampoline'' mechanism with underlying Lyapunov structure, our work identifies gating-induced stabilization of ghost-like transients---accompanied by emergent multistability---as a physical mechanism underlying extended retrieval without assuming a Lyapunov function.

\begin{acknowledgments}
We thank Matthew Du, Dmitry Krotov, Peter B. Littlewood, Rama Ranganathan, and Greg J. Stephens for insightful discussions.
D.G. further acknowledges fruitful discussions with Robert L. Jack during the early stages of this work, and support from the Yoshida Scholarship Foundation. H.M.L.R. and S.V. acknowledge support from the National Institute of
General Medical Sciences of the NIH under Award No.~R35GM147400. 
D.G., H.M.L.R., M.S., and S.V. also acknowledge support from the NSF through the Physics Frontier Center for Living Systems (Grant No.~PHY-2317138). M.S. is funded by the Max-Planck Society which supported this work.
H.M.L.R. is additionally supported by a postdoctoral fellowship from the University of Chicago Materials Research Science and Engineering Center, which is funded by the NSF under Award No.~DMR-2011854.
This work was completed in part with resources provided by the University of Chicago’s Research Computing Center.
\end{acknowledgments}

\bibliography{ref_hopfield}

\begin{thebibliography}{40}
\bibitem [{Not()}]{NoteTimeDependentStability}%
  \BibitemOpen
  \href@noop {} {}\bibinfo {note} {For linear systems with time-dependent
  coefficients, instantaneous eigenvalues alone do not determine asymptotic
  stability; for example, a system may be asymptotically stable even when its
  instantaneous coefficient matrix has an eigenvalue with positive real
  part.}\BibitemShut {Stop}%
\end{thebibliography}

\newpage


\clearpage
\onecolumngrid
\pagenumbering{arabic}
\vspace{1em}

\setcounter{equation}{0}
\setcounter{figure}{0}
\setcounter{table}{0}
\setcounter{section}{0}
\setcounter{subsection}{0}

\renewcommand{\theequation}{S\arabic{equation}}
\renewcommand{\thefigure}{S\arabic{figure}}
\renewcommand{\thetable}{S\arabic{table}}

\renewcommand{\theHequation}{supp.equation.\arabic{equation}}
\renewcommand{\theHfigure}{supp.figure.\arabic{figure}}
\renewcommand{\theHtable}{supp.table.\arabic{table}}
\renewcommand{\theHsection}{supp.section.\arabic{section}}
\renewcommand{\theHsubsection}
  {supp.section.\arabic{section}.\arabic{subsection}}


\makeatletter
\begingroup
\frontmatter@setup

\frontmatter@title@above
{%
  \frontmatter@title@format
  Supplementary Information for:\\[0.2cm]
  ``Neuromodulation--inspired gated associative memory networks:\\
  extended memory retrieval and emergent multistability''
  \par
}
\frontmatter@title@below

{%
  \normalfont\normalsize
  \frontmatter@authorformat
  \mbox{Daiki Goto\textsuperscript{\S,1,2}},
  \enspace
  \mbox{Hector Manuel Lopez Rios\textsuperscript{\S,2,3}},
  \enspace
  \mbox{Monika Scholz\textsuperscript{4}},
  \enspace
  \mbox{Suriyanarayanan Vaikuntanathan\textsuperscript{2,3}}
  \par
}

\frontmatter@above@affiliation@script
{%
  \normalfont
  \frontmatter@affiliationfont
  \textsuperscript{\normalfont 1}Department of Physics,
  The University of Chicago, Chicago, Illinois 60637, USA\\
  \textsuperscript{\normalfont 2}The James Franck Institute,
  The University of Chicago, Chicago, Illinois 60637, USA\\
  \textsuperscript{\normalfont 3}Department of Chemistry,
  The University of Chicago, Chicago, Illinois 60637, USA\\
  \textsuperscript{\normalfont 4}Max Planck Institute for Neurobiology of
  Behavior—caesar, Bonn 53175, Germany
  \par
}

\frontmatter@authorbelow
\endgroup
\makeatother

\maketitle

\onecolumngrid
\thispagestyle{empty} 

\setcounter{equation}{0}
\setcounter{figure}{0}
\setcounter{table}{0}
\renewcommand{\theequation}{S\arabic{equation}}
\renewcommand{\thefigure}{S\arabic{figure}}
\renewcommand{\thetable}{S\arabic{table}}

\setcounter{section}{0}
\setcounter{subsection}{0}
\setcounter{subsubsection}{0}
\makeatletter


\renewcommand{\thesection}{S\arabic{section}}
\renewcommand{\thesubsection}{\thesection.\arabic{subsection}}
\renewcommand{\thesubsubsection}{\thesubsection.\arabic{subsubsection}}
\def\p@subsection{}
\def\p@subsubsection{}

\makeatother

\phantomsection
\color{\editcolor}
\section*{Contents}
\let\ORIGaddcontentsline\addcontentsline   
\renewcommand{\addcontentsline}[3]{%
  \ifstrequal{#1}{toc}
    {\ifstrequal{#2}{section}{\ORIGaddcontentsline{#1}{#2}{#3}}{}}
    {\ORIGaddcontentsline{#1}{#2}{#3}}}
\startcontents[si]
{\hypersetup{linkcolor=ruri}
\printcontents[si]{}{1}{}}
\vspace{1em}
\color{black}

\begingroup

\renewcommand{\citenumfont}[1]{S#1}
\renewcommand{\bibnumfmt}[1]{[S#1]}

\begin{bibunit}[apsrev4-2]

\makeatletter
\@startbibunitorrelax
\if@filesw
  \immediate\write\@bibunitaux{\string\citation{SIcontrol}}
\fi
\makeatother

\section{Illustration of realistic neuromodulator networks}
The multilayer connectome \cite{bentley2016multilayer} was updated using recent receptor expression and neuropeptide connectome datasets. The synaptic connectome was extracted from \cite{Cook2019Wholeanimal} and the bilateral pairs were merged to allow comparison with the transcriptome data, which cannot discriminate bilateral neurons. The expression levels for the dopamine and serotonin receptors were extracted from the single cell transcriptome dataset for hermaphrodites (CenGen) with an expression threshold = 2. Neuropeptide connectomes were extracted from \cite{ripoll_neuropep_celegans} using the mid-range model and three arbitrarily chosen ligand-receptor pairs. 


\section{Methods}
Unless otherwise stated, we use random iid binary patterns, $\bm{\xi}^\mu\sim\mathsf{Unif}(\{-1,1\})$, and the system size was set to $N=1000$.
The dynamics defined in Eqs.\ts(\ref{eq:hebb}--\ref{eq:gRNN_z}) were integrated using the Euler method with a fixed time step $dt$.
\par
For Fig.\ts\ref{fig:main}, we used $dt=0.2$.
The network was initialized as $\mathbf{x}(0)=c_1\bm{\xi}^1+\bm{\eta}$, where $\bm{\eta}\sim\mathcal{N}(\bm{0},I_N)$ and the coefficient $c_1$ was determined numerically to satisfy the target initial overlap $m(0)$. 
The neuromodulatory units were initialized as $\mathbf{z}(0)\sim\mathcal{N}(\bm{0},I_N)$.
To characterize the steady-state quantities, we simulated the dynamics for a duration $T=2000$ (corresponding to $T/dt$ steps), which was verified to be sufficient for the system {\color{\editcolor} of size $N=1000$} (for both ungated $\gamma=0$ and binary gated $\gamma\to\infty$ networks) to relax to a steady state. 
Steady-state overlaps were averaged over $n_\text{avg}=100$ independent realizations of the patterns $\{\xi_i^\mu\}$ and initial conditions $\{\mathbf{x}(0),\mathbf{z}(0)\}$.
Fig.\ts\ref{fig:transient_retrieval} and the relevant panels of Fig.\ts\ref{fig:dmft-mft}(a)--(c) used the same parameters, with $n_\text{avg}=30$ and $n_\text{avg}=500$, respectively. 
For the phase flow analysis in Fig.\ts\ref{fig:phase_flow}, we used a two-pattern initialization $\mathbf{x}(0)=c_1\bm{\xi}^1+c_2\bm{\xi}^2$ with varying $c_1,c_2$ and a time step $dt=0.25$.

\subsection{Numerical integration of DMFT}
The numerical integration of the DMFT equations for producing Fig.\ts\ref{fig:dmft-mft}(a)--(c) was performed using a custom code adapted from the open-source implementation by Clark~\cite{clark2025transientdynamicsassociativememory} available at \url{https://github.com/davidclark1/TransientDynamicsAssocMem}.
For optimization, we also referred to the one by Roy \emph{et al.}~\cite{Roy-Biroli-Bunin-Cammarota19} available at \url{https://github.com/FelixRoy/dmft_rLV}. 
We modified the original solver to incorporate the self-adaptive gating dynamics and the associated auxiliary variables.
\par
Key numerical parameters used in the integration include the number of DMFT trajectories (population size) $M$, the time step $dt$, and the number of self-consistency iterations.
For the curves in Fig.\ts\ref{fig:dmft-mft}(a)--(c), we used $M=10^4$ and $dt=0.02$ and iterated the self-consistency loop until the macroscopic observables (in particular $m(t)$ and $C_\phi(t,t')$) converged within a small fixed tolerance.
To improve numerical stability, the order parameters were updated using linear mixing (with a reinjection rate or a ``memory factor'' $\zeta_\text{m}$; see also Sec.\ts\ref{sec:asymptotic_dmft} for detail) as in Refs.~\cite{Roy-Biroli-Bunin-Cammarota19,clark2025transientdynamicsassociativememory}.

\section{Dynamical mean-field theory }
\label{sec:si_dmft}
In this section, we provide details of the dynamical mean-field theory (DMFT) for the gated associative memory network defined by Eqs.\ts(\ref{eq:hebb}--\ref{eq:gRNN_z}).
For many-body dynamical systems in the thermodynamic limit, DMFT derives an effective low-dimensional dynamics in which the complex many-body correlations are encoded in the non-Markovianity of the process.
We refer the reader to recent pedagogical introductions to DMFT~\cite{Zou-Huang24:rev, galla2024generatingfunctionalanalysisrandomlotkavolterra} and a recent survey~\cite{Cugliandolo24:rev}.
\par
Our derivation of the DMFT equations is based on the dynamical cavity method \cite{Mezard-Parisi-Virasoro86,Mezard-Parisi-Virasoro87:book}, that was originally invented to study the Sherrington--Kirkpatrick model of spin glasses, whose central idea lies in analyzing the response of a high-dimensional system to the addition of a few extra degrees of freedom, based on linear response theory \cite{Kubo57} and the central limit theorem. 
We specifically follow the recently developed bipartite dynamical cavity approach~\cite{clark2025transientdynamicsassociativememory}.
Alternative approaches using the path-integral formalism~\cite{Crisanti-Sompolinsky18} for related systems are detailed, for example, in Refs.\ts\cite{Krishnamurthy-Can-Schwab22,Aguilera-Morales-Rosas-Shimazaki25,mimura2025dynamicalpropertiesdenseassociative,kabashima2025dynamicalmeanfieldapproach}. 
\par
Without loss of generality, we consider the retrieval of a single condensed pattern $\bm{\xi}^1$; the remaining patterns (including those introduced by the cavity method) are treated as uncondensed patterns.
The extension of this analysis to multiple condensed patterns is straightforward.

\subsection{Derivation of self-consistent DMFT equations via the cavity method}
We begin by recasting the equations of motion (\ref{eq:hebb}--\ref{eq:gRNN_z}) in the main text in the following form:
\begin{align}
    \partial_tx_i &= \sigma(z_i)\bigg[ -x_i+\frac{g}{\sqrt{\alpha}}\bigg(\sum_{\mu=1}^P\xi_i^\mu m_\mu + h_i^x\bigg) \bigg]   \label{eq:dxdt}
    \\
    \tau_z\partial_tz_i &= - z_i + \frac{1}{\sqrt{N}}\sum_{j=1}^NW_{ij}\phi_j + h_i^z \ , \label{eq:dzdt}
\end{align}
where $\phi_i=\phi(x_i)$ and $m_\mu(t)$ is the overlap between the pattern $\bm{\xi}^\mu$ and the activated output of the system $\{\phi_i\}_{i=1}^N$ at time $t$ defined by
\begin{equation}\label{eq:m_with_b}
    m_\mu(t) = \frac{1}{N}\sum_{i=1}^N\xi_i^\mu\phi_i + b_\mu \ .
\end{equation}
In Eqs.\ts(\ref{eq:dxdt}--\ref{eq:m_with_b}), the external fields $h^x_i(t),h^z_i(t)$ and the bias $b_\mu(t)$ are introduced solely to probe the response to the extra `cavity' degrees of freedom added in the subsequent derivation.
\par
To implement the bipartite cavity method, we now add test cavity degrees of freedom $x_0(t), z_0(t)$ along with a cavity uncondensed pattern indexed $\mu=0$.
Specifically, we introduce the pattern entries $\{\xi_0^\mu\}_{\mu=0}^P$ (the new neuron's couplings to all patterns) and $\{\xi_i^0\}_{i=1}^N$ (the new pattern's values across the existing neurons); this expands the system to $2(N+1)$ state variables and $P+1$ stored patterns (see Fig.\ts\ref{fig:cavity_schematic}).
\begin{figure*}[b]
    \centering
    \includegraphics[width=0.68\textwidth]{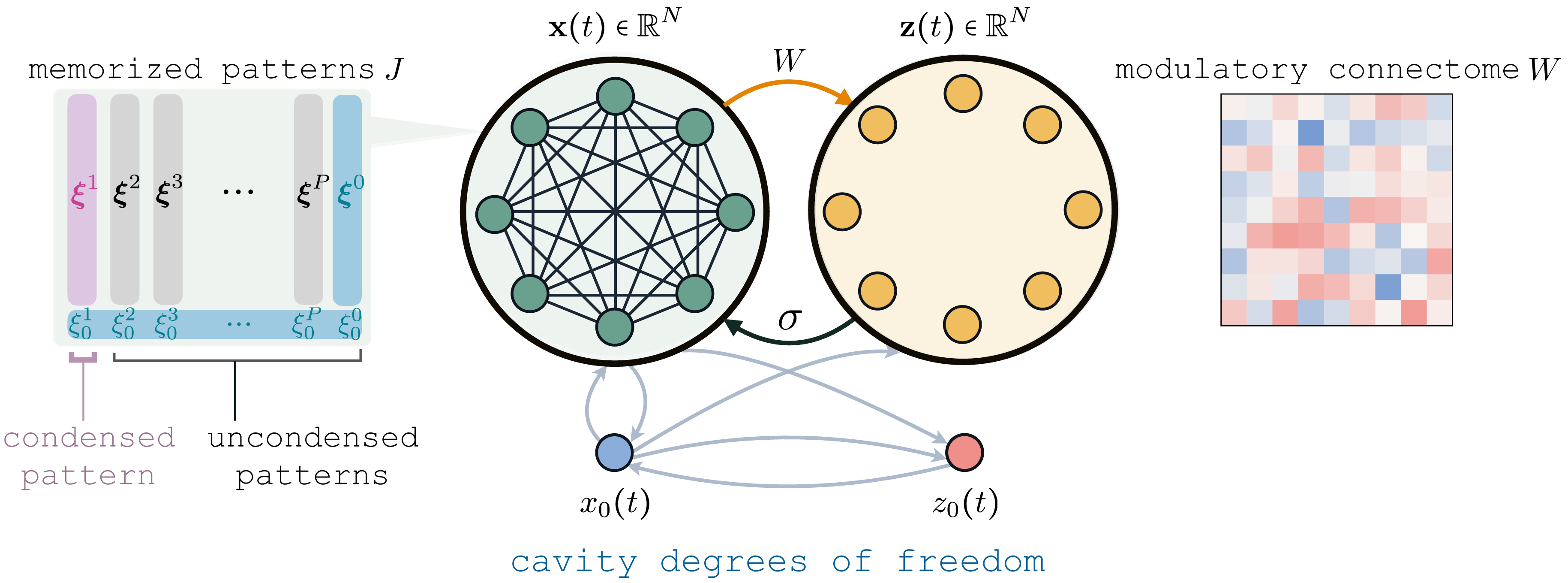}
    \caption{ 
    \label{fig:cavity_schematic}
    Schematic of the gated associative memory network analyzed by the cavity method.
    }
\end{figure*}

Naïvely, the effects of these added cavities are expected to be small (at most $\mathcal{O}(1/\sqrt{N})$) in the thermodynamic limit ($N, P \to \infty$ with $\alpha=P/N$ remaining of order one). 
Consequently, their inclusion can be treated perturbatively using the linear response theory:
\begin{align}
    \delta{\phi}_i(t) &=  \sum_{j} \xi_j^0 \int \! dt' \, \chi_{ij}(t,t') m_0(t') \ ,
    \\
    \delta{z}_i(t) &= \frac{1}{\sqrt{N}}\sum_{j} W_{j0} \int \! dt' \, \rho_{ij}(t,t') \phi_0(t') \ ,
    \\
    \delta{m}_\mu(t) &= \frac{1}{N} \sum_{\nu} \xi_0^\nu \int \! dt' \, \gamma_{\mu\nu}(t,t') \phi_0(t') \ ,
\end{align}
where we have defined the response functions (susceptibilities) $\chi_{ij}(t,t')=\frac{\delta \phi_i(t)}{\delta h_j^x(t')}$, $\rho_{ij}(t,t')=\frac{\delta{z}_i(t)}{\delta{h}^z_j(t')}$, and $\gamma_{\mu\nu}(t,t')=\frac{\delta{m}_\mu(t)}{\delta{b}_\nu(t')}$.
Causality implies that responses are retarded, meaning $\chi_{ij}(t,t') = \rho_{ij}(t,t') = \gamma_{\mu\nu}(t,t') = 0$ for all $t < t'$.
We note that the variables $\phi_i$ and $z_i$, and the condensed overlap $m_1$, are $\mathcal{O}(1)$, whereas the uncondensed overlaps $m_\mu$, $\mu\ne1$, are $\mathcal{O}(N^{-1/2})$.
\par
Substituting these perturbations back into the equation of motion for the cavity variable $x_0$, we obtain
\begin{align}
    \partial_t{x}_0 &= \sigma(z_0)\bigg[ -x_0 + \frac{g}{\sqrt{\alpha}}\sum_{\mu=0}^P \xi_0^\mu (m_\mu + \delta{m}_\mu) \bigg]
    \nonumber\\
    &= \sigma(z_0)\bigg[ -x_0+\frac{g}{\sqrt{\alpha}}\bigg( 
    \underbrace{ \xi_0^1 m_1 }_{\text{signal}}
    +\textcolor{kaba}{ \underbrace{ \sum_{\mu(\neq1)}\xi_0^\mu m_\mu }_{\text{crosstalk noise} \ \eta_0^x(t)} }
    + \textcolor{ruri}{ \underbrace{\frac{1}{N}\sum_{\mu,\nu}\xi_0^\mu\xi_0^\nu \int \! dt' \, \gamma_{\mu\nu}(t,t')\phi_0(t') }_{\text{Onsager reaction term}}}
    \bigg) \bigg] \ . \label{eq:dx0dt_expanded}
\end{align}
Here, we have separated the contribution of the single condensed pattern $(\mu=1)$---the \textit{signal term}---from the \textcolor{kaba}{crosstalk noise} (due to uncondensed patterns) and the \textcolor{ruri}{Onsager reaction term} (due to the cavity response).
[Recall that $m_1 \sim \mathcal{O}(1)$ while $m_{\mu(\neq{1})} \sim \mathcal{O}(1/\sqrt{N})$.]
The \textcolor{kaba}{crosstalk noise} and \textcolor{ruri}{Onsager reaction term} in Eq.\ts\eqref{eq:dx0dt_expanded} are stochastic functions of time, whose statistical properties can be determined in the thermodynamic limit via the central limit theorem.
\par
The \textcolor{kaba}{crosstalk noise $\eta_0^x(t)$} represents the noise whose first moment is zero (since $\langle \xi_0^\mu \rangle = 0$) and whose second moment is calculated as
\begin{align}
    \sum_{\mu,\nu \neq 1} \langle \xi_0^\mu\xi_0^\nu \rangle \langle m_\mu(t) m_\nu(t') \rangle 
    &= \sum_{\mu \neq 1} \langle m_\mu(t) m_\mu(t') \rangle 
    \approx P \langle m_0(t) m_0(t') \rangle \ ,
\end{align}
where we used the independence of patterns $\langle \xi_0^\mu\xi_0^\nu \rangle = \delta_{\mu\nu}$ and the fact that all uncondensed patterns are statistically equivalent.
\par
On the other hand, the mean of the \textcolor{ruri}{Onsager reaction term} is given by
\begin{align}
    \frac{1}{N}\sum_{\mu,\nu} \langle \xi_0^\mu\xi_0^\nu \rangle \int dt' \langle \gamma_{\mu\nu}(t,t') \rangle \phi_0(t')
    &= \frac{1}{N} \sum_{\mu} \int \! dt' \, \langle \gamma_{\mu\mu}(t,t') \rangle \phi_0(t') \nonumber \\
    &= {\alpha}\int dt' \, G(t,t')\phi_0(t') \ ,
\end{align}
where we have defined the population-averaged response function $G(t,t')=\frac{1}{P}\sum_\mu \langle \gamma_{\mu\mu}(t,t') \rangle$.
This term yields the leading-order contribution, while fluctuations around the mean are subleading in the thermodynamic limit.
\par
Thus, to sum up, we arrive at an effective single-site process (dropping the cavity index $i=0$):
\begin{align}
    \partial_t x &= \sigma(z)\bigg[ -x + \frac{g}{\sqrt{\alpha}}\bigg( \xi^1 m_1 + \eta_x  + {\alpha}\int \! dt' \, G(t,t')\phi(x(t')) \bigg) \bigg] \ ,
    \label{eq:dx0dt_final}
    \\
    &\langle{\eta_x(t)}\rangle = 0 \ , \quad \langle{\eta_x(t)\eta_x(t')}\rangle = P \langle{m_0(t)m_0(t')}\rangle \ , \label{eq:etax_var0}
\end{align}
in which the many-body correlations of the original dynamics are now encoded in the \emph{non-Markovianity} of the dynamics.
Note that $m_0(t)$ here represents the effective overlap of an uncondensed pattern; thus $\langle m_0(t)m_0(t') \rangle = \mathcal{O}(1/N)$ and accordingly $\langle{\eta_x(t)\eta_x(t')}\rangle = \mathcal{O}(\alpha)=\mathcal{O}(1)$.
\par
Applying similar manipulations to the auxiliary variable dynamics $\tau_z\partial_tz=-z+(1/\sqrt{N})\sum_i W_{0i}(\phi_i+\delta\phi_i)$ (where $\delta\phi_i=\phi'_i\delta{x}_i$ with the prime denoting the derivative) and the overlap for the uncondensed pattern yields
\begin{align}
    \tau_z\partial_t{z} &= - z + \eta_z(t) \ ,\label{eq:dz0dt_final}
    \\
    m_0(t) &= \zeta(t) + \int dt' \, R(t,t')m_0(t') + b_0(t) \ , \label{eq:m0} 
\end{align}
where we have defined the population-averaged local response function $R(t,t') = \frac{1}{N}\sum_i \langle \chi_{ii}(t,t') \rangle$, and the effective noises $\eta_z(t)$ and $\zeta(t)$ satisfy
\begin{equation}
    \langle{\eta_z(t)}\rangle = 0 \ , \quad \langle{\eta_z(t)\eta_z(t')}\rangle = C_\phi(t,t') 
\end{equation}
and
\begin{equation}
    \langle{\zeta(t)}\rangle = 0 \ , \quad \langle{\zeta(t)\zeta(t')}\rangle = \frac{1}{N} C_\phi(t,t') \ ,
    \label{eq:zeta}
\end{equation}
respectively.
\par
Equations\ts(\ref{eq:dx0dt_final}--\ref{eq:zeta}), together with the conditions closing the self-consistency loop
\begin{align}
    C_\phi(t,t') &= \langle{\phi(t)\phi(t')}\rangle \ , \label{eq:C_phi} \\
    R(t,t') &= \bigg\langle\frac{\delta{\phi}(t)}{\delta{h}^x(t')}\bigg\rangle \bigg|_{h^x=0} \ , \label{eq:R_tt} \\
    G(t,t') &= \bigg\langle\frac{\delta m_0(t)}{\delta b_0(t')}\bigg\rangle \bigg|_{b_0=0} \ ,
\end{align}
collectively define the DMFT of the gated associative memory network. 
Here, the averages $\langle \dots \rangle$ denote the expectation values over the realizations of the effective stochastic process (i.e., over the noises $\{\eta_x, \eta_z, \zeta\}$, the pattern variable $\xi^1$, and initial configurations $\{{x}(0),z(0)\}$).

We can further simplify the set of equations by integrating out the overlap for the uncondensed pattern $m_0$.
To this end, we first take the derivative of Eq.\ts\eqref{eq:m0} with respect to the bias $b_0(t')$ and then take the average to obtain 
\begin{equation}
    G(t,t') = \delta(t-t') + \int{du}\,R(t,u)G(u,t') \ ,
\end{equation}
where the Dirac $\delta$ term represents a temporally local perturbation.
This recursive relation can be viewed as a Dyson equation between the bare propagator $\delta$, the self-energy kernel $R$, and the dressed propagator (or Green function) $G$~\cite{Dyson49,Kamenev11:book}. 
Denoting the latter formally by
\begin{equation}
    G(t,t') = (\delta-R)^{-1}(t,t') \ ,
\end{equation}
one can write the solution to Eq.\ts\eqref{eq:m0} formally as
\begin{equation}
    m_0(t) = \int\!{dt'}\,G(t,t')\zeta(t') \ ,
\end{equation}
where $b_0\to0$ has been taken. 
Using this, one can then rewrite the noise variance \eqref{eq:etax_var0} as
\begin{align}
    \langle{\eta_x(t)\eta_x(t')}\rangle &= 
        P\int\!{du}{du'}\,G(t,u)\langle{\zeta(u)\zeta(u')}\rangle G(t',u') \ .
\end{align}
Substituting Eq.\ts\eqref{eq:zeta} further yields
\begin{equation}
    \langle{\eta_x(t)\eta_x(t')}\rangle = \alpha \int\!{dudu'}\,G(t,u)C_\phi(u,u')G(t',u') \ . 
\end{equation}

\subsection{Off-diagonal response is sub-dominant}
In deriving the DMFT equations, we implicitly assumed that off-diagonal susceptibilities are subleading compared to their diagonal counterparts in the thermodynamic limit, positing the ansatz: 
\begin{align}\label{eq:ansatz_chi}
    \chi_{ij}(t,t') = \begin{cases}
        \mathcal{O}(1) \ , & \text{for $i=j$}\\
        \mathcal{O}(1/\sqrt{N}) \ , & \text{for $i\neq j$} 
    \end{cases} \ .
\end{align}
Here, following similar arguments presented, \emph{e.g.}, in Refs.\ts\cite{Roy-Biroli-Bunin-Cammarota19, Prueser-Rosmej-Engel24}, we briefly demonstrate that this ansatz is self-consistent with the given many-body dynamics \eqref{eq:dxdt}.
\par
Taking the functional derivative of Eq.\ts\eqref{eq:dxdt} with respect to $h^x_j(t')$, we obtain, for $t>t'$,
\begin{align}\label{eq:dchi_micro}
    \partial_t\tilde{\chi}_{ij}(t,t') 
    &= \sigma(z_i)\bigg[-\tilde{\chi}_{ij}(t,t')
    +\underbrace{\frac{g}{\sqrt{\alpha}N}\sum_k\sum_\mu\xi_i^\mu\xi_k^\mu\chi_{kj}(t,t')}_{=\vcentcolon S_{ij}(t,t')}  \bigg] \ ,
\end{align}
where $\tilde{\chi}_{ij}(t,t')\vcentcolon=\chi_{ij}(t,t')/\phi_i'(t)$.
Up to an $\mathcal{O}(1)$ prefactor, the second term in parentheses (denoted $S_{ij}$) can be rewritten using the Hebbian coupling $\tilde{J}_{ik}=\frac{1}{P}\sum_\mu\xi_i^\mu\xi_k^\mu$ as
\begin{align}\label{eq:S_ij}
    S_{ij}(t,t') = \mathcal{O}(1)\sum_k \tilde{J}_{ik}\chi_{kj}(t,t') \ .
\end{align}
Assuming that the pattern distribution $P_\xi$, from which the patterns are drawn iid, has zero mean and unit variance, $\tilde{J}_{ii}$ has mean $1$.
For off-diagonal terms ($k\neq i$), $\tilde{J}_{ik}$ has zero mean and variance $1/P$.
Conequently, the sum in Eq.\ts\eqref{eq:S_ij} contains a `signal' contribution from $k=i$ (proportional to $\chi_{ij}$) and a zero-mean `noise' contribution from the remaining $k\neq{i}$ terms. 
Summing $N-1$ such independent terms yields net fluctuations that scale as $\mathcal{O}(1/\sqrt{N})$.
\par
Therefore, we find that $S_{ii}=\mathcal{O}(1)$, whereas for $i\neq j$, $S_{ij}=\mathcal{O}(1/\sqrt{N})$.
Due to causality, the dynamics governed by Eq.\ts\eqref{eq:dchi_micro} preserves this scaling for all $t>t'$, rendering the result self-consistent with the ansatz \eqref{eq:ansatz_chi}.
\par
A similar discussion holds for the other response functions $\rho_{ij}(t,t')$ and $ \gamma_{\mu\nu}(t,t')$.

\subsection{Time-asymptotic DMFT equations}\label{sec:asymptotic_dmft}
We obtain the stationary DMFT equations by setting Eqs.\ts\eqref{eq:dx0dt_final} and \ts\eqref{eq:dz0dt_final} to zero and removing any time dependency in Eqs.\ts\eqref{eq:C_phi} \&\ts\eqref{eq:R_tt}. 
Using simple algebra, we obtain the following time-asymptotic DMFT equations for gated RNNs:  
\begin{align}
    \sigma(z^*) x^* &= \sigma(z^*) \left [ \frac{g}{\sqrt{\alpha}}(\xi m^* + \eta_x) + g\sqrt{\alpha}K \phi \left ( x^* \right ) \right ], \label{eq:x_ss} \\
    z^* &= \eta_z, \\
    m^* &= \left < \xi \phi \left ( x^* \right) \right >, \label{eq:m_ss}\\
    C_\phi^* &=  \langle \phi \left ( x^* \right )^2  \rangle, \label{eq:C_ss}\\
    R^* &= \left < \frac{ (g/\sqrt{\alpha})\sigma(z^*) \phi'(x^*)}{ \sigma(z^*) \left[ 1 - g\sqrt{\alpha}K \phi'(x^*) \right] }\right >, \label{eq:R_ss}\\
    K &= \frac{1}{1-R^*}.
\end{align}
These equations describe the time-asymptotic statistical properties of the cavity neuron that was inserted during our DMFT formulation. Similarly, we solve this set of equations self-consistently for the order parameters defined by Eqs.\ts\eqref{eq:m_ss}--\ts\eqref{eq:R_ss}. Each iteration consists of sampling $\xi$, $\eta_x$, and $\eta_z$ from their respective distributions. 
The stationary Gaussian noises have zero mean and variances $\langle\eta_x^2\rangle=\alpha K^2C_\phi^*$ and $\langle\eta_z^2\rangle=C_\phi^*$.
Specifically, an iteration consists of solving for $x^\ast$ using Eq.\ts\eqref{eq:x_ss}, and calculating and updating the order parameters by a memory factor $\zeta_\text{m}$~\cite{Roy-Biroli-Bunin-Cammarota19,clark2025transientdynamicsassociativememory}, $$X = \zeta_\text{m} \,X_\mathrm{old} + (1-\zeta_\text{m}) \, X_\mathrm{new}, $$
where $X\in\{ m^\ast,C_\phi^\ast, R^\ast \}$ represents an order parameter. 
Additionally, cavity neurons are naturally partitioned into active and closed neurons given the explicit presence of the gating function $\sigma(z^\ast)$ in Eq.\ts\eqref{eq:x_ss}. To account for the correct statistics of the closed single-site neurons, we extracted their approximate steady state overlap $m^\ast_\mathcal{F}$ and correlation $C^\ast_{\phi_\mathcal{F}}$ values from DMFT calculations. Due to the computational complexity of our DMFT computations, approximate steady state values of these order parameters were obtained at $t=50$. 
Furthermore, we calculate populations of closed single-site neurons solving a modified version of Eq.\ts\eqref{eq:x_ss} such that it satisfies the approximate steady state closed neuron order parameter values obtained from DMFT calculations. 
Finally, we consider closed neurons to have a null contribution to the susceptibility $R^\ast_\mathcal{F} = 0$. 

Using these methods we constructed Fig.\ts\ref{fig:dmft-mft}(d) by only probing the upper and lower bounds of the time-asymptotic overlap band at different values of the memory load $\alpha$ using different initial overlap conditions. We calculated the upper and lower bounds of the time-asymptotic overlap at each $\alpha$ by obtaining $m^\ast_{\mathcal{F}}$ and $C^\ast_{\phi_\mathcal{F}}$ from DMFT calculations at $t = 50$ using $m(0) = 1.0$ and $m(0) = 0.035$, respectively. 

\color{\editcolor}
\subsection{Non-Gaussianity and probing the thermodynamic limit} 
\label{sec:si_dmft_nongauss}

In associative memory networks with the symmetric Hebbian coupling, the field generated by the non-condensed patterns is not Gaussian~\cite{nishimori1993retrieval,ozeki1994noise}, and there is no reason to expect otherwise in a gated network. Since the DMFT of Sec.\ts\ref{sec:dmft} replaces this field by an effective Gaussian noise together with a retarded response, we must check that the resulting description reproduces the field the network actually experiences. We therefore track the skewness and excess kurtosis of the non-condensed field over time, in many-body simulations at several system sizes and using DMFT. In simulations the field is measured directly as $$h_i^\mathrm{sim}(t) = \frac{g}{\sqrt{\alpha}}\sum_{\mu=2}^{P} \xi_i^\mu m^\mu(t),$$  while in the DMFT it is the sum of the effective noise and the retarded self-interaction,  $$h^\mathrm{DMFT}(t) = \frac{g}{\sqrt{\alpha}}\left(\eta_x(t) + \alpha\!\int\! dt'\, G(t,t')\,\phi(x(t'))\right).$$ The two are the same physical quantity, but only the DMFT expression separates it into noise and response; that decomposition is not observable at finite $N$, so we compare the total field. 
\par
Figure~\ref{fig:dmft_summary}(a),(b) shows the two moments. The skewness vanishes for the larger networks and in the DMFT trajectory, whereas the excess kurtosis is negative at all times, growing in magnitude from $\simeq-0.13$ at early times to $\simeq-0.55$ by the end of the integrated window. The many-body curves converge onto the DMFT trajectory with increasing $N$, so the theory reproduces the field distribution including its non-Gaussian part. The theory attributes this departure from Gaussian noise to the retarded response, since the effective noise $\eta_x$ is Gaussian by construction. This attribution is consistent with the observed time dependence: $\phi = \tanh(x)$ is bounded and saturates as the dynamics settle, which thins the tails of the total field and accounts for the growth of $|{\rm excess\ kurtosis}|$ from $0.13$ to $0.55$ over the window. The decomposition into noise and response is not separately observable at finite $N$, so we do not test the Gaussianity of $\eta_x$ in isolation.
\par
Figure~\ref{fig:dmft_summary}(c),(d) shows the corresponding convergence of the order parameters: the condensed overlap $m(t)$ and the equal-time correlation $C_\phi(t,t)$. These converge more quickly than the field moments: the order parameters are already close to the DMFT prediction by $N \simeq 500$, whereas the fourth moment of the field requires $N \gtrsim 1000$, as expected of a higher-order statistic.
\begin{figure*}[!htbp]
    \centering
    \includegraphics[width=0.8\textwidth]{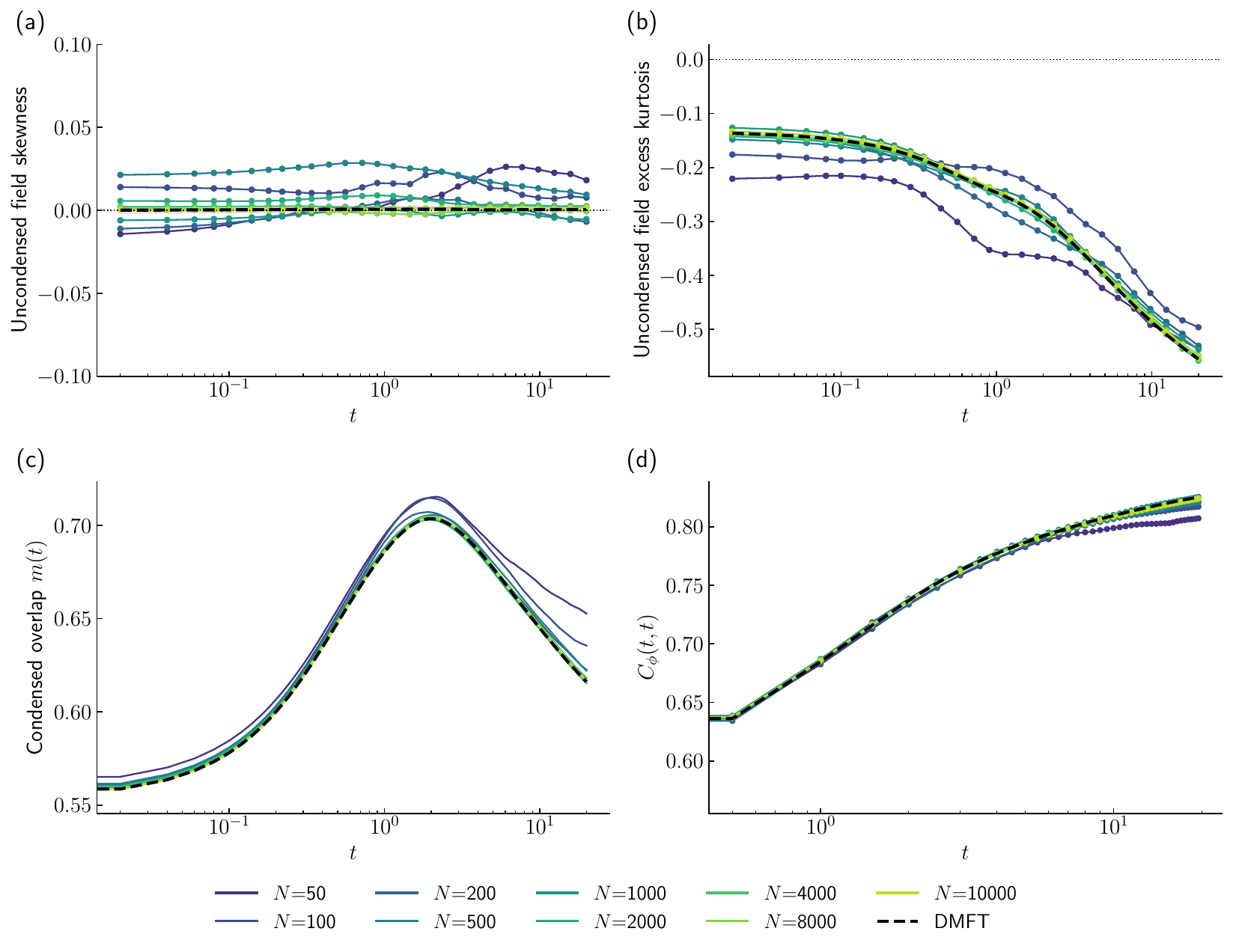}
    \caption{{\color{\editcolor}\textbf{Finite-$N$ simulations converge to the DMFT, including the non-Gaussian statistics of the non-condensed field.} (a) Skewness and (b) excess kurtosis of the non-condensed field $h_i(t)$, as a function of time, for gated networks $(\gamma\to\infty)$ of increasing size (colours) and using DMFT (black dashed). The field takes one value per neuron, so each point is a moment of the distribution across neurons and disorder realizations at that time. (c) Condensed overlap $m(t)$ and (d) equal-time correlation $C_\phi(t,t)$ for the same simulations. Zero excess kurtosis would indicate Gaussian tails; the negative values reflect the bounded retarded contribution to the field, which the DMFT computes explicitly. Parameters: $\alpha = 0.4, g = 1.5, m(0) \simeq 0.56, dt = 0.02$, averaged over 100 disorder realizations per system size; the DMFT solution uses $M = 10000$ sample trajectories.}
    }
    \label{fig:dmft_summary}
\end{figure*}
\color{black}

\color{\editcolor}
\clearpage
\section{Frozen-stabilization and Jacobian spectra}
\label{sec:jacobian-spectra}
In this section, we analyze the instantaneous Jacobian of the gated dynamics and show that the stabilizing mechanism at work in our gated Hebbian network is an associative memory realization of the \textit{frozen-stabilization} proposed by Can and Krishnamurthy~\cite{Can-Krishnamurthy25} for gated random recurrent networks. We first derive the exact spectral decomposition of the Jacobian in the hard-gate limit and identify the characteristic spectral signature of frozen-stabilization, which our network reproduces during retrieval [Fig.\ts\ref{fig:spectra}(a)]. We then discuss what selects the retrieved state among the stabilized ones, and finally treat finite gate steepness and the ungated limit [Fig.\ts\ref{fig:spectra}(b,c)].
\par\medskip
For later convenience, we rewrite the full gated recurrent dynamics as
\begin{align}
  \partial_t{\mathbf{x}} &= D_{{\sigma}}\mathbf{f}_\mathrm{RNN}(\mathbf{x}) \ , 
  \label{eq:gRNN-x-compact}
  \\
  \tau_z\partial_t{\mathbf{z}} &= -\mathbf{z}+\bar{W}\phi(\mathbf{x}) \ ,
  \label{eq:gRNN-z-compact}
\end{align}
where we define
\begin{equation}
  \mathbf{f}_\mathrm{RNN}(\mathbf{x})=-\mathbf{x}+\bar{J}\phi(\mathbf{x}) \ ,
  \qquad
  \bar{J}=\frac{g}{\sqrt{PN}}J \ ,
  \qquad
  \bar{W}=\frac{1}{\sqrt N}W \ ,
\end{equation}
and the diagonal matrices
\begin{equation}
  D_{\sigma}=\diag[\sigma(\mathbf{z})] \ ,
  \qquad
  D_{\phi^\prime}=\diag[\phi'(\mathbf{x})] \ ,
  \qquad
  D_{\sigma^\prime f}=\diag[\sigma'(\mathbf{z})\odot \mathbf{f}_\mathrm{RNN}(\mathbf{x})] \ .
\end{equation}
Here and below, $\sigma$ and $\phi$ act componentwise, and $\odot$ denotes the elementwise Hadamard multiplication. 

Linearizing the dynamics (\ref{eq:gRNN-x-compact}, \ref{eq:gRNN-z-compact}) about the instantaneous state $(\mathbf{x}(t),\mathbf{z}(t))$ gives the $2N\times2N$ instantaneous Jacobian
\begin{equation}
    \mathcal{J}(t) =
    \begin{pmatrix}
        D_\sigma(-I_N+\bar{J}D_{\phi'}) & D_{\sigma'f} \\[2pt]
        \tau_z^{-1}\bar{W}D_{\phi'} & -\tau_z^{-1}I_N
    \end{pmatrix} \ .
    \label{eq:general-jacobian}
\end{equation}
The two off-diagonal blocks form the $\mathbf{x}\to\mathbf{z}\to\mathbf{x}$ feedback loop. This feedback via $D_{\sigma'f}\bar{W}D_{\phi'}$ vanishes in the hard-gate limit. In contrast, at finite $\gamma$, it is generally nonzero during a transient and leads to more nontrivial $\bar{W}$-mediated relaxation dynamics.

\subsection{Hard-gate limit: exact spectral decomposition}
In the hard-gate limit $\gamma\to\infty$, the gate becomes binary, $\sigma(z_i>0)=1$ and $\sigma(z_i<0)=0$. At a regular time for which $z_i(t)\neq0$ for every $i$, we define the active and the frozen subpopulations
\begin{equation}
    \mathcal{A}_t = \{i \,|\, z_i(t)>0 \} \ ,
    \qquad\mathcal{F}_t = \{i \,|\, z_i(t)<0 \}
\end{equation}
with 
\begin{equation}
    N_{\mathcal{A}_t} = |\mathcal{A}_t| \ ,
    \qquad
    N_{\mathcal{F}_t} = |\mathcal{F}_t| \ ,
    \qquad
    N_{\mathcal{A}_t}+N_{\mathcal{F}_t}=N
\end{equation}
Note that, at an exact switching time, one or more $z_i(t)=0$ and the ordinary instantaneous Jacobian is not well-defined; the analysis below therefore applies piecewise, on intervals over which the active-frozen partition is fixed, and we suppress the time subscript within such an interval. Away from switching surfaces, $\sigma'(z_i)=0$ for every $i$, so $D_{\sigma'f}=0$ in Eq.\ts\eqref{eq:general-jacobian}, while $D_\sigma$ reduces to the indicator of $\mathcal{A}$.

Denoting $\mathbf{x}_\mathcal{F}=\{{x}_i\,|\,i\in\mathcal{F}_t\}$ and $\mathbf{x}_\mathcal{A}=\{{x}_i\,|\,i\in\mathcal{A}_t\}$, the dynamics can be reordered in the form
\begin{align}
  \partial_t{\mathbf{x}}_{\mathcal F}
  &=0 \ ,
  \label{eq:hard-frozen-dynamics}
  \\
  \partial_t{\mathbf{x}}_{\mathcal A}
  &=
  \underbrace{-\mathbf{x}_{\mathcal A}+\bar{J}_{\mathcal A}\phi(\mathbf{x}_{\mathcal A})}_{\text{active (dynamic)}}
  +\underbrace{\bar{J}_{\mathcal F}\phi(\mathbf{x}_{\mathcal F})}_{\text{static field}\:\:\mathbf{b}_\mathcal{F}} \ ,
  \label{eq:hard-active-dynamics}
  \\
  \tau_z\partial_t{\mathbf z}
  &=
  -\mathbf z+\bar W_{\mathcal A}
  \phi(\mathbf{x}_{\mathcal A})+\bar W_{\mathcal F}\phi(\mathbf{x}_{\mathcal F}) \ ,
  \label{eq:hard-z-dynamics}
\end{align}
where $\bar{J}_{\mathcal A}\in\mathbb{R}^{N_{\mathcal A}\times N_{\mathcal A}}$,
$\bar{J}_{\mathcal F}\in\mathbb{R}^{N_{\mathcal A}\times N_{\mathcal F}}$,
$\bar W_{\mathcal A}\in\mathbb{R}^{N\times N_{\mathcal A}}$, and
$\bar W_{\mathcal F}\in\mathbb{R}^{N\times N_{\mathcal F}}$
denote the corresponding subblocks of $\bar J$ and $\bar W$
associated with the active and frozen coordinates. Conditioned on the frozen values $\mathbf x_{\mathcal{F}}$, the active population thus evolves as a recurrent network driven by the static input/field
\begin{equation}
    \mathbf{b}_\mathcal{F} \vcentcolon= \bar{J}_{\mathcal F}\phi(\mathbf{x}_{\mathcal F}) \ .
\end{equation}
This is the defining dynamical motif of frozen-stabilization~\cite{Can-Krishnamurthy25}: a subpopulation ceases to evolve and supplies a static background field to the rest. Depending on its strength and structure, such a field can shift active units away from the high-gain region of the transfer function and thereby stabilize their dynamics, as shown for chaotic random recurrent networks~\cite{Schuecker-Goedeke-Helias18,Ogawa-Fumarola-Mazzucato23} and, more relevantly here, for memory-retrieval dynamics under input-correlated external fields~\cite{Engel-Englisch-Schutte89,Amit-Parisi-Nicolis90}.

Defining $D_{\phi'_\mathcal A}=\diag[\phi'(\mathbf{x}_\mathcal{A})]$, $D_{\phi'_\mathcal F}=
  \diag[\phi'(\mathbf{x}_\mathcal{F})]$,
and the Jacobian of the active population
\begin{equation}
    L_\mathcal{A} \vcentcolon= -I_{N_\mathcal{A}}+\bar{J}_{\mathcal{A}}D_{\phi'_\mathcal{A}} \ ,
\end{equation}
the corresponding instantaneous Jacobian, in the reordered coordinate $(\mathbf{x}_\mathcal{F},\mathbf{x}_\mathcal{A},\mathbf{z})$, is given by
\begin{equation}
    \mathcal{J} = \begin{pmatrix}
        0 & 0 & 0 \\
        \bar{J}_{\mathcal{F}}D_{\phi'_\mathcal{F}} & L_\mathcal{A} & 0 \\
        \tau_z^{-1}\bar{W}_{\mathcal{F}}D_{\phi'_\mathcal{F}} & \tau_z^{-1}\bar{W}_{\mathcal{A}}D_{\phi'_\mathcal{A}} & -\tau_z^{-1}I_N
    \end{pmatrix} \ ,
    \label{eq:hard-full-jacobian}
\end{equation}
where $0$-blocks have the dimensions implied by the active-frozen partition.

Because Eq.\ts\eqref{eq:hard-full-jacobian} is block lower triangular, its characteristic polynomial factorizes exactly as 
\begin{equation}
    \det[\lambda I_{2N}-\mathcal{J}] = \lambda^{N_{\mathcal{F}_t}}(\lambda+1/\tau_z)^N\det[\lambda I_{N_\mathcal{A}}-L_{\mathcal{A}}] \ ,
\end{equation}
and the instantaneous spectrum therefore separates into three sectors:
\begin{equation}
    \spec({\mathcal{J}}) = \{0\}^{N_{\mathcal{F}_t}} \,\sqcup\,\{-1/\tau_z\}^{N} \,\sqcup\, \spec{(L_{\mathcal{A}})} \ ,
    \label{eq:hard-spectrum}
\end{equation}
where $\sqcup$ denotes the disjoint union. 
For the symmetric Hebbian coupling considered here, the active spectrum $\spec(L_{\mathcal{A}})$ is real (see SI Sec.\ts\ref{sec:interlacing-stabilization}).

The interpretation of the active sector $\spec(L_\mathcal{A})$ depends on the state at which $L_\mathcal{A}$ is evaluated.
Consider first a stationary state, $\partial_t\mathbf{x}_{\mathcal{A}}=\partial_t\mathbf{z}=0$. At such a state, $L_{\mathcal{A}}$ is time-independent, and its eigenvalues determine local stability transverse to the frozen directions exactly. The state is not asymptotically stable in the full phase space, however, because its spectrum contains $N_{\mathcal{F}}$ exact zero marginal modes. A sufficiently small displacement of $\mathbf{x}_{\mathcal{F}}$ is not restored; instead, the active and modulatory variables relax to nearby stationary values (while the active-frozen partition remains unchanged). Provided that $L_{\mathcal{A}}$ is nonsingular, these nearby states form an $N_{\mathcal{F}}$-dimensional sheet parameterized locally by $\mathbf{x}_{\mathcal{F}}$, with the frozen zero modes tangent to this sheet. The appropriate stability criterion is therefore
\begin{equation}
  \lambda_\perp
  \vcentcolon=\max_{\lambda\in\spec(L_{\mathcal A})}\RePart\lambda<0 \ ,
  \label{eq:transverse-stability}
\end{equation}
which guarantees decay of perturbations transverse to the sheet; we call such states \emph{transversely stable}. 
If $L_{\mathcal A}$ is itself singular, additional zero modes appear beyond the $N_\mathcal{F}$ frozen modes.

During a transient, in contrast, both $L_{\mathcal{A}}(t)$ and the active-frozen partition evolve, so the instantaneous spectrum serves only as a diagnostic of the stabilization mechanism~\cite{NoteTimeDependentStability}, as suggested in Ref.~\cite{Krishnamurthy-Can-Schwab22}. The characteristic spectral signature of frozen-stabilization is still the coexistence of an extensive exact zero-mode sector, tangent to the sheet, and a negative transverse edge $\lambda_\perp$; Fig.\ts\ref{fig:frozenstab_schematic} illustrates this structure schematically. In Fig.\ts\ref{fig:spectra}(a), the motion of $\lambda_\perp(t)$ into the negative half-line diagnoses the development of frozen-stabilization along the transient, while the final spectrum, evaluated at the stationary state, satisfies Eq.\ts\eqref{eq:transverse-stability} exactly.
\begin{figure*}[!htbp]
    \centering
    \includegraphics[width=0.34\textwidth]{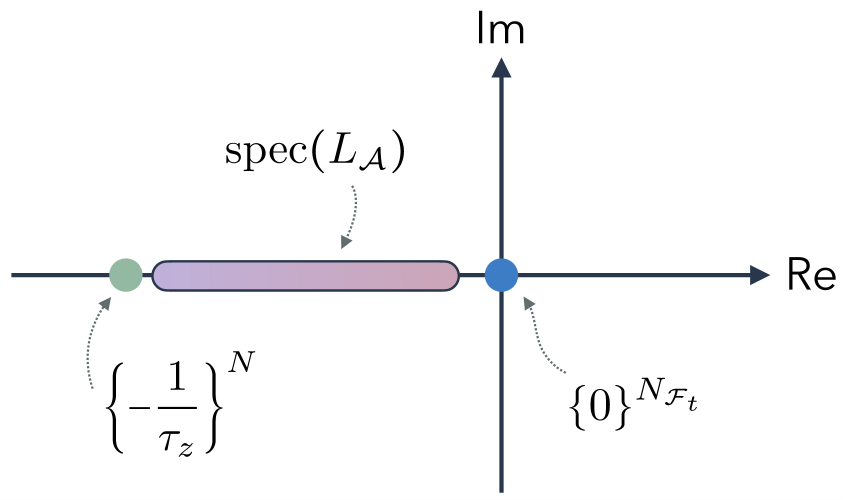}
    \caption{{\color{\editcolor}\textbf{Schematic of the frozen-stabilization spectral signature.} }}
    \label{fig:frozenstab_schematic}
\end{figure*}

Lastly, we note that, although $\bar{W}$ does not appear explicitly in the instantaneous eigenvalues \eqref{eq:hard-spectrum}, it implicitly determines how the active-frozen partition dynamics evolves, and hence which principal block $L_\mathcal{A}$ and which frozen field $\mathbf{b}_\mathcal{F}$ are realized. For the Hebbian coupling, this field is pattern-correlated by construction:
\begin{equation}
    \mathbf{b}_{\mathcal{F}} = \frac{g}{\sqrt{\alpha}}\sum_{\mu=1}^{P}\bm{\xi}^\mu_{\mathcal{A}}m^\mu_{\mathcal{F}} \ ,
    \qquad
    m^\mu_{\mathcal{F}} = \frac{1}{N}\sum_{j\in\mathcal{F}}\xi_j^\mu\,\phi(x_j) \ ,
    \label{eq:bF-pattern}
\end{equation}
i.e., a coherent component along the cued pattern (say, $\mu=1$), weighted by the frozen partial overlap $m^1_{\mathcal{F}}$, plus incoherent crosstalk from the uncondensed patterns $(\mu\geq2)$. During retrieval, $m^{1}_{\mathcal{F}}=\rho_{\mathcal{F}_t}m_{\mathcal{F}_t}$ remains of order one [cf.\ the closed-population overlap dynamics $m_{\mathcal{F}_t}(t)$ in the inset of main-text Fig.\ts\ref{fig:transient_retrieval}(a), and its steady-state values parameterizing the band of main-text Fig.\ts\ref{fig:dmft-mft}(d)], so the active subsystem is driven by a self-consistently determined field correlated with the cued pattern---a setting in which such a field, when imposed externally, was shown to stabilize retrieval beyond the classical capacity~\cite{Engel-Englisch-Schutte89,Amit-Parisi-Nicolis90}---and the activity-dependent gating thereby selects frozen states whose field supports the queried memory. 

\begin{figure*}[!htbp]
    \centering
    \includegraphics[width=0.98\textwidth]{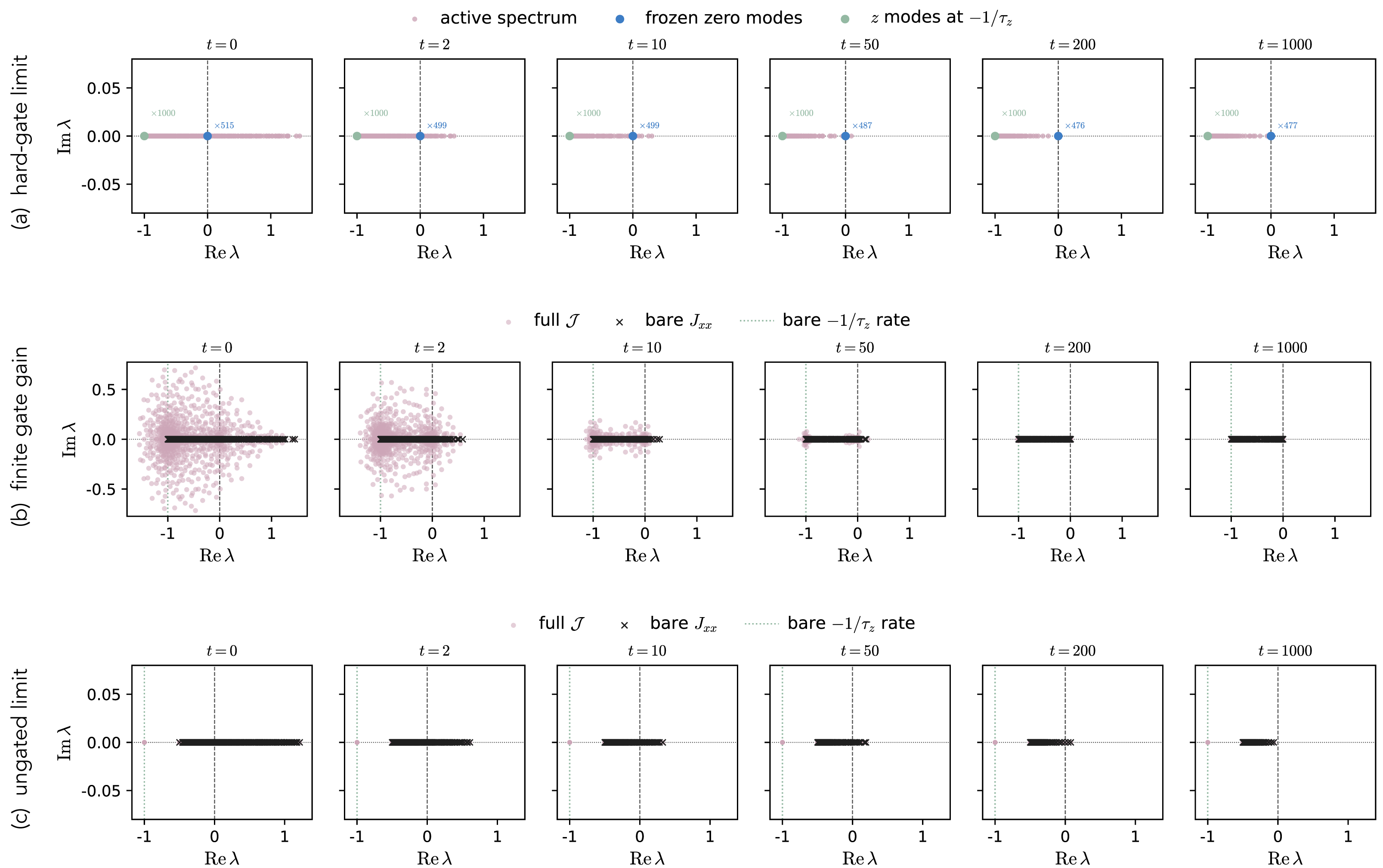}
    \caption{\color{\editcolor} \textbf{Evolution of the instantaneous Jacobian spectrum during retrieval.}
    Spectra are shown at representative times for a single disorder realization.
    \textbf{(a)} In the hard-gate limit, $\gamma\to\infty$, the spectrum separates, between gate-switching events, into the active neuronal spectrum (pink), one frozen zero mode per closed coordinate (blue), and the $N$ modulatory relaxation modes at $-1/\tau_z$ (green). The active spectrum is real for the symmetric Hebbian coupling, and its right edge moves into the stable half-line, while the frozen zero modes remain marginal, displaying the characteristic spectral signature of frozen stabilization.
    \textbf{(b)} At finite gate gain, $\gamma=10$, the bare neuronal block $J_{xx}$ (black crosses) has a real spectrum, whereas the full Jacobian $\mathcal J$ (pink) develops complex-conjugate eigenvalues during the transient through the $W$-mediated $\mathbf{x}\to\mathbf{z}\to\mathbf{x}$ feedback. The long transient is accompanied by many near-zero slow modes; as the system approaches a true fixed point, the feedback term $D_{\sigma' f}\bar W D_{\phi'}$ vanishes and the full spectrum returns to the real axis.
    \textbf{(c)} In the ungated limit, $\gamma=0$, the $\mathbf{x}\to\mathbf{z}\to\mathbf{x}$ feedback vanishes and the Jacobian is block triangular, so the instantaneous spectrum remains real.
    Parameters: $N=1000$, $\alpha=0.4$, $m(0)=0.55$, and $\tau_z=1$.
    \label{fig:spectra}
    }
\end{figure*}

\subsection{Stabilizing tendency of freezing for symmetric recurrent coupling}
\label{sec:interlacing-stabilization}
For the symmetric Hebbian coupling considered here, one can show that freezing has a one-sided stabilizing tendency (not destabilizing). Since $\phi'>0$, the active Jacobian $L_{\mathcal{A}}$ is similar to the symmetric matrix
\begin{equation}
    H_{\mathcal{A}}=-I_{N_{\mathcal{A}}}+D_{\phi'_{\mathcal{A}}}^{1/2}\bar{J}_{\mathcal{A}}D_{\phi'_{\mathcal{A}}}^{1/2} \ ,
\end{equation}
and therefore has a real spectrum. Moreover, $H_{\mathcal{A}}$ is a principal submatrix of the symmetric representation of the full neuronal Jacobian,
\begin{equation}
    H=-I_N+D_{\phi'}^{1/2}\bar{J}D_{\phi'}^{1/2} \ .
\end{equation}
The Cauchy interlacing theorem~\cite{fisk2005shortproofcauchysinterlace} then gives
\begin{equation}
    \lambda_{\max}(L_{\mathcal{A}})=\lambda_{\max}(H_{\mathcal{A}})\leq\lambda_{\max}(H) \ .
\end{equation}
Thus, at any given neural configuration $\mathbf{x}$, restricting the dynamics to the active coordinates cannot increase the leading spectral edge relative to the full neuronal Jacobian evaluated at the same $\mathbf{x}$.
Interlacing does not guarantee $\lambda_\perp<0$, because the reduced edge may remain positive and the frozen field also changes the state and its local gains. This one-sided result is specific to symmetric $\bar{J}$; for a general asymmetric recurrent coupling, interlacing does not apply and freezing may either stabilize or destabilize the active dynamics.

\subsection{Finite gate steepness and the ungated limit}
\label{sec:finite-gamma-jacobian}
At finite $\gamma$, no coordinate is exactly frozen. 
Although the neuronal block ${J}_{xx}=D_\sigma(-I_N+\bar{J}D_{\phi'})$ has a real spectrum for symmetric $\bar{J}$, the feedback $D_{\sigma'f}\bar{W}D_{\phi'}$ can generate complex-conjugate eigenvalues during a transient [Fig.\ts\ref{fig:spectra}(b)]. The exact hard-gate zero modes are replaced by near-zero slow modes that underlie the prolonged retrieval transient dynamics.

At a true finite-$\gamma$ fixed point, $\sigma(z_i)>0$ implies $\mathbf{f}_\mathrm{RNN}(\mathbf{x}^*)=0$. Consequently, $D_{\sigma'f}^*=0$ and the feedback vanishes, and the Jacobian becomes block lower triangular so that
\begin{equation}
    \spec(\mathcal{J}^*) = \spec(J_{xx}^*) \,\sqcup\, \{-1/\tau_z\}^{N} \ .
\end{equation}
For symmetric Hebbian coupling $\bar{J}$, the neuronal Jacobian $J_{xx}^*$ is similar to
\begin{equation}
H_\gamma
=(D_\sigma^*)^{1/2}H^*(D_\sigma^*)^{1/2} \ ,
\qquad
H^*=-I_N+(D_{\phi'}^*)^{1/2}\bar{J}(D_{\phi'}^*)^{1/2} \ .
\end{equation}
The matrices $H_\gamma$ and $H^*$ are congruent, so Sylvester's law of inertia~\cite{Sylvester1852} implies that they have the same numbers of positive, negative, and zero eigenvalues. Finite gating thus changes relaxation and escape rates but not the stability classification of a fixed point. Together with the fixed-point correspondence established in SI Sec.\ts\ref{sec:order-of-limits}, this shows that the finite-$\gamma$ retrieval plateau is a long-lived quasi-frozen transient rather than a new attractor.

In the ungated limit $\gamma=0$, $\sigma\equiv1/2$ and $\sigma'\equiv0$, so $D_{\sigma'f}=0$ throughout the trajectory. The feedback is therefore absent, the Jacobian remains block triangular, and its instantaneous spectrum remains real at all times, eventually reaching a fixed point corresponding to the cue-dependent remanent overlap~\cite{Gardner-Derrida-Mottishaw87} [see Fig.\ts\ref{fig:spectra}(c)].

\subsection{New fixed-point manifolds can arise only in the hard-gate limit}
\label{sec:order-of-limits}
For every finite gate steepness $\gamma<\infty$, the logistic gate is strictly positive, $\sigma(z_i)>0$. Any fixed point $(\mathbf{x}^*,\mathbf{z}^*)$ of the gated dynamics therefore satisfies
\begin{equation}
    D_\sigma^*\,\mathbf{f}_\mathrm{RNN}(\mathbf{x}^*)=0
    \;\Longrightarrow\;
    \mathbf{f}_\mathrm{RNN}(\mathbf{x}^*)=0 \ ,
\label{eq:fp-implication}
\end{equation}
together with $\mathbf{z}^*=\bar{W}\phi(\mathbf{x}^*)$, so the fixed points of the full dynamics are in one-to-one correspondence with those of the ungated Amari--Hopfield network. The parameters $W$, $\tau_z$, and $\gamma$ reshape the relaxation trajectories and their timescales, but cannot create new $\mathbf{x}$-fixed points. 

In the hard-gate limit $\gamma\to\infty$, in contrast, some gates vanish exactly, $\sigma(z_i^*)=0$, and Eq.\ts\eqref{eq:fp-implication} no longer holds. A hard-gate fixed state need only satisfy $[f_{\mathrm{RNN},i}(\mathbf{x}^*)]_{i\in\mathcal{A}}
={0}$,
while $[f_{\mathrm{RNN},i}(\mathbf{x}^*)]_{i\in\mathcal{F}}$
may remain nonzero. New fixed points, exact zero modes, and continuous fixed-point manifolds are therefore genuine hard-gate phenomena.

\clearpage
\section{Robustness to modulatory dynamics and implementation}
\label{sec:modulatory-robustness}

\subsection{Dependence on the modulatory timescale and initial gate state}
\label{sec:tauz-initialization}
To determine whether hard-gate ($\gamma\to\infty)$ retrieval remains accessible when the modulatory variables evolve substantially more slowly than the neural variables, here we vary the ratio of modulatory to neural timescales ($\tau_x$, which is implicitly set to be unity) over more than three decades for $\tau_z\in\{0.1,0.3,1,3,10,30,100,300\}$. 

We compared two initial conditions for
the modulatory variables:
\begin{equation}
z_i(0)\sim\mathcal{N}(0,1)
\quad\text{and}\quad
z_i(0)=1 \ .
\end{equation}
The former (the one used in the main text) initially closes approximately half of the binary gates, whereas the latter initially leaves every neural coordinate active. For each disorder realization, the same $\{\bm{\xi}^\mu\}_{\mu=1}^P, W$, and initial cue $\mathbf{x}(0)$ were used for all values of $\tau_z$ and for both modulatory initializations. Results were averaged over
eight independent disorder realizations. To secure the convergence of the trajectory, we used the maximum integration times $4500\tau_x$ for $\tau_z<300$ and $9000\tau_x$ for $\tau_z=300$.
The corresponding ungated network was relaxed using
the same patterns and neural cues, and is shown as the gray dashed reference.
\begin{figure*}[b]
    \centering
    \includegraphics[width=0.99\textwidth]{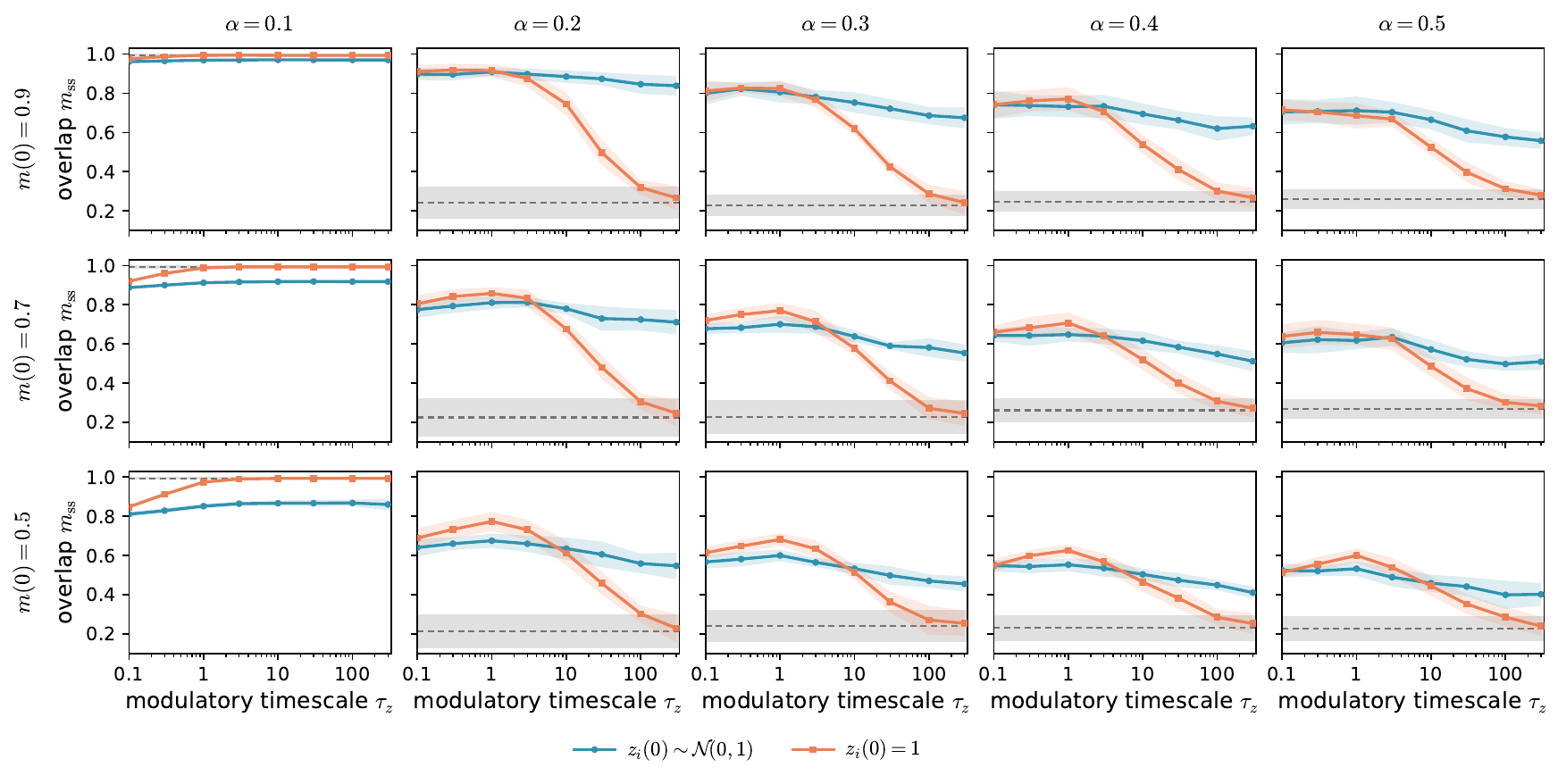}
    \caption{{\color{\editcolor}\textbf{Dependence of hard-gate retrieval on the modulatory timescale and gate initialization.} Steady-state overlap $m_{\mathrm{ss}}$ vs. modulatory time-constant $\tau_z\in\{0.1,0.3,1,3,10,30,100,300\}$. Columns correspond to memory loads $\alpha=0.1,0.2,0.3,0.4,0.5$, while rows correspond, from top to bottom, to initial cue overlaps $m(0)=0.9,0.7,0.5$. Blue circles denote random modulatory initial conditions, $z_i(0)\sim\mathcal{N}(0,1)$ (used in the main text), and orange squares denote an initially open gate configuration, $z_i(0)=1$. Solid curves show means over eight independent disorder realizations, and shaded regions show one standard deviation. The gray dashed line and shaded band show the corresponding mean and standard deviation of the ungated network. Within each realization, all values of $\tau_z$ and both modulatory initializations use the same patterns, coupling matrix, and cue noise. Parameters used: $N=1000$, $g=1.5$, $\gamma\to\infty$, and $n_\text{avg}=8$.}}
    \label{fig:tauz-initialization}
\end{figure*}

Figure~\ref{fig:tauz-initialization} shows that slow modulatory dynamics do not by themselves eliminate high-overlap hard-gated states, but their accessibility depends strongly on the initial gate configuration. For $z_i(0)\sim\mathcal{N}(0,1)$, an extensive subset of coordinates is closed immediately, and an order-one overlap persists even through the largest tested value, $\tau_z=300$, including at loads well above the conventional retrieval capacity. In contrast, when all gates are initially open, the overlap is comparable at small $\tau_z$ but generally decreases as $\tau_z$ becomes large, approaching the ungated low-overlap state at high load. In this case, the network first evolves approximately under the ungated dynamics, and gate closure occurs only after the slowly evolving modulatory fields change sign to negative. 

These results are consistent with the $\tau_z$-independence of the fixed-point equations: $\tau_z$ and the initial conditions determine which hard-gated steady state is selected, rather than which steady states exist. Together with the $\mathbf{x}(0)$-dependence discussed in the main text, the dependence on $\mathbf{z}(0)$ provides an additional means of selecting persistent hard-gated states, potentially useful for input- or context-driven computation.

\subsection{Dependence on the form of the modulatory coupling $W$}
\label{sec:W-robustness}
The iid modulatory matrix $W_{ij}\sim\mathcal{N}(0,1)$ used in the main text is the maximally unstructured baseline, chosen to disentangle the effect of heterogeneous multiplicative gating from any designed modulatory architecture. In biological networks, however, the effective coupling is plausibly low-rank or sparse. Here we test whether gated retrieval depends on these properties of $W$; Figures~\ref{fig:lowrank-dynamics} and~\ref{fig:sparse-dynamics} show the resulting overlap dynamics across gate steepnesses $\gamma$, initial overlaps $m(0)$, and several decades of rank ratio $R/N$ and connection probability $p$.

For low-rank $W$, we generate factorized couplings
\begin{equation}
    W = \frac{1}{\sqrt{N}}\sum_{\rho=1}^{R}\mathbf u^\rho(\mathbf v^\rho)^{\top} \ ,
    \qquad
    \mathbf u^\rho,\mathbf v^\rho \sim\mathcal N(\bm 0,I_N) \ ,
\end{equation}
with all factors independent, so that $\operatorname{rank}W=R$ almost surely while $\langle W_{ij}\rangle_W=0$, $\langle W_{ij}^2\rangle_W=R/N$, and $\langle W_{ij}W_{ji}\rangle_W=0$ for $i\neq j$. The case $R=N$ is a full-rank factorized ensemble rather than a strictly iid one.

For sparse $W$, we dilute the coupling as
\begin{equation}
    W_{ij}=B_{ij}G_{ij} \ ,
    \qquad
    B_{ij}\sim\operatorname{Bernoulli}(p) \ ,
    \qquad
    G_{ij}\sim\mathcal N(0,1) \ ,
\end{equation}
with all variables independent, so that $\langle W_{ij}^{2}\rangle_W=p$ and $p=1$ recovers the dense iid ensemble.

In both Figures~\ref{fig:lowrank-dynamics} and~\ref{fig:sparse-dynamics}, the extreme values $R=0$ and $p=0$ give $W=0$. In the hard-gate limit (rightmost columns), retrieval is largely robust to rank and density, with the high-overlap state persisting for sufficiently strong cues across the tested range. At large but finite gate steepness, by contrast, reducing $R$ or $p$ shortens the retrieval plateau. Both trends reflect the scale of the modulatory field: the typical magnitude of $z_i$ grows with ${\smash[b]{\langle W_{ij}^2\rangle}}_W^{1/2}$, so the product $\gamma\,|z_i|$ acts as an effective gate steepness, of order $\gamma\sqrt{R/N}$ and $\gamma\sqrt{p}$, which is lowered by reducing rank or density at fixed finite $\gamma$ --- whereas the binary gate depends only on $\operatorname{sign}(z_i)$ and is therefore more insensitive to the overall scale of $W$. Hard-gate retrieval thus does not require $W$ to be dense or full rank, consistent with the reported robustness of frozen-stabilization to the structure of the gating interactions~\cite{Can-Krishnamurthy25}.

\begin{figure*}[!hbtp]
    \centering
    \includegraphics[width=0.8543\textwidth]{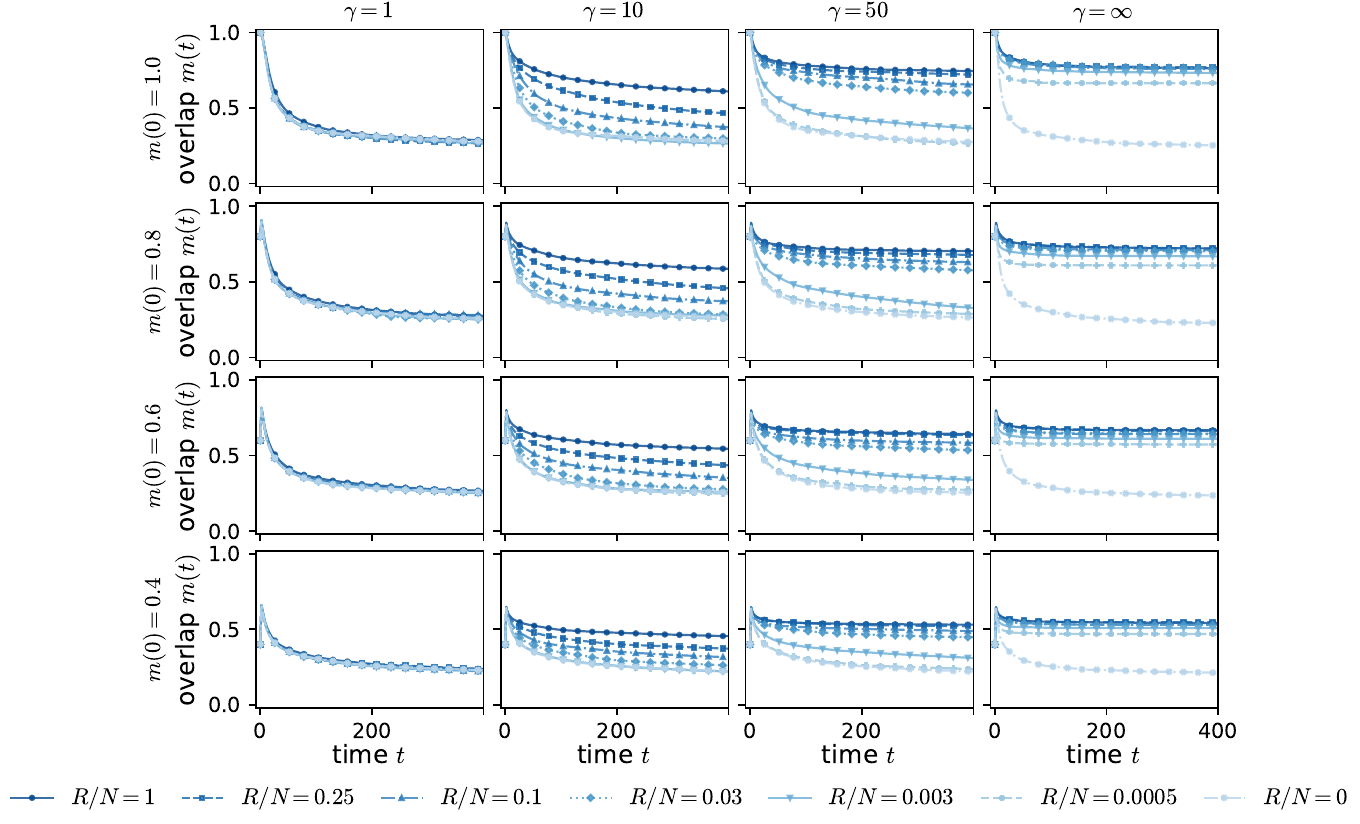}
    \caption{{\color{\editcolor}\textbf{Overlap dynamics under low-rank modulatory coupling.}
    Rows correspond, from top to bottom, to initial overlaps $m(0)=1.0,\,0.8,\,0.6,\,0.4$, while columns correspond, from left to right, to gate steepnesses $\gamma=1,\,10,\,50,$ and $\infty$. Within each panel, curves show rank ratios $R/N=1,\,0.25,\,0.1,\,0.03,\,0.003,\,0.0005,$ and $0$, ordered from darkest to lightest blue. The low-rank modulatory matrix is set by $W=(1/\sqrt{N})\sum_{\rho=1}^R\mathbf{u}_\rho\mathbf{v}_\rho^\top$ with independent Gaussian vectors $\mathbf{u}_\rho,\mathbf{v}_\rho\sim\mathcal{N}(\mathbf{0},I_N)$.
    Parameters: $N=2000$, $\alpha=0.4$, $\tau_z=1$, $z_i(0)=1$, and $n_{\text{avg}}=12$.}}
    \label{fig:lowrank-dynamics}
\end{figure*}

\begin{figure*}[!hbtp]
    \centering
    \includegraphics[width=0.7\textwidth]{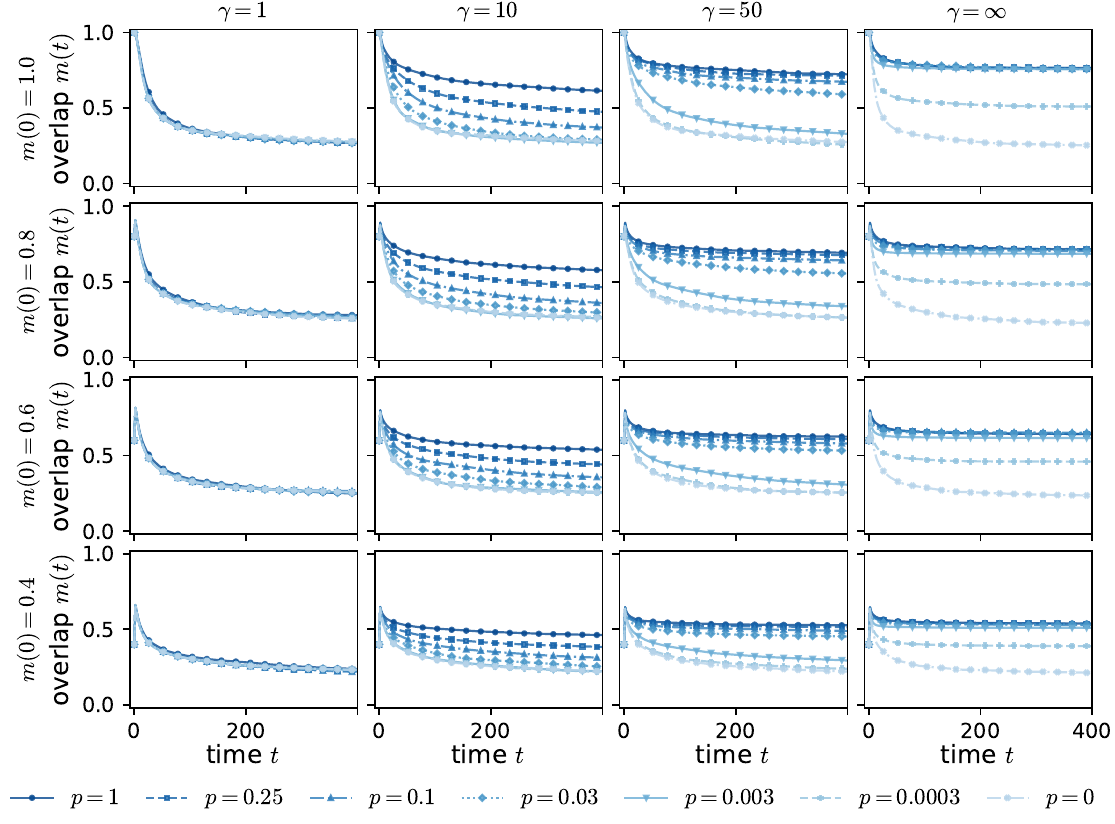}
    \caption{{\color{\editcolor}\textbf{Overlap dynamics under sparse modulatory coupling.}
    Rows correspond, from top to bottom, to initial overlaps $m(0)=1.0,\,0.8,\,0.6,\,0.4$, while columns correspond, from left to right, to gate steepnesses $\gamma=1,\,10,\,50,$ and $\infty$. Within each panel, curves show connection probabilities $p=1,\,0.25,\,0.1,\,0.03,\,0.003,\,0.0003,$ and $0$, ordered from darkest to lightest blue, where $p=1$ corresponds to the dense limit used in the main text. The sparse modulatory matrix is set by $W_{ij}=B_{ij}G_{ij}$, where $B_{ij}\sim\operatorname{Bernoulli}(p)$ and $G_{ij}\sim\mathcal{N}(0,1)$.
    Parameters: $N=2000$, $\alpha=0.4$, $\tau_z=1$, $z_i(0)= 1$, and $n_{\text{avg}}=12$.}}
    \label{fig:sparse-dynamics}
\end{figure*}

\subsection{Dependence on the construction of initial cue $\mathbf{x}(0)$}
\label{sec:cue-construction}

\begin{figure*}[!htbp]
    \centering
    \includegraphics[width=0.9\textwidth]{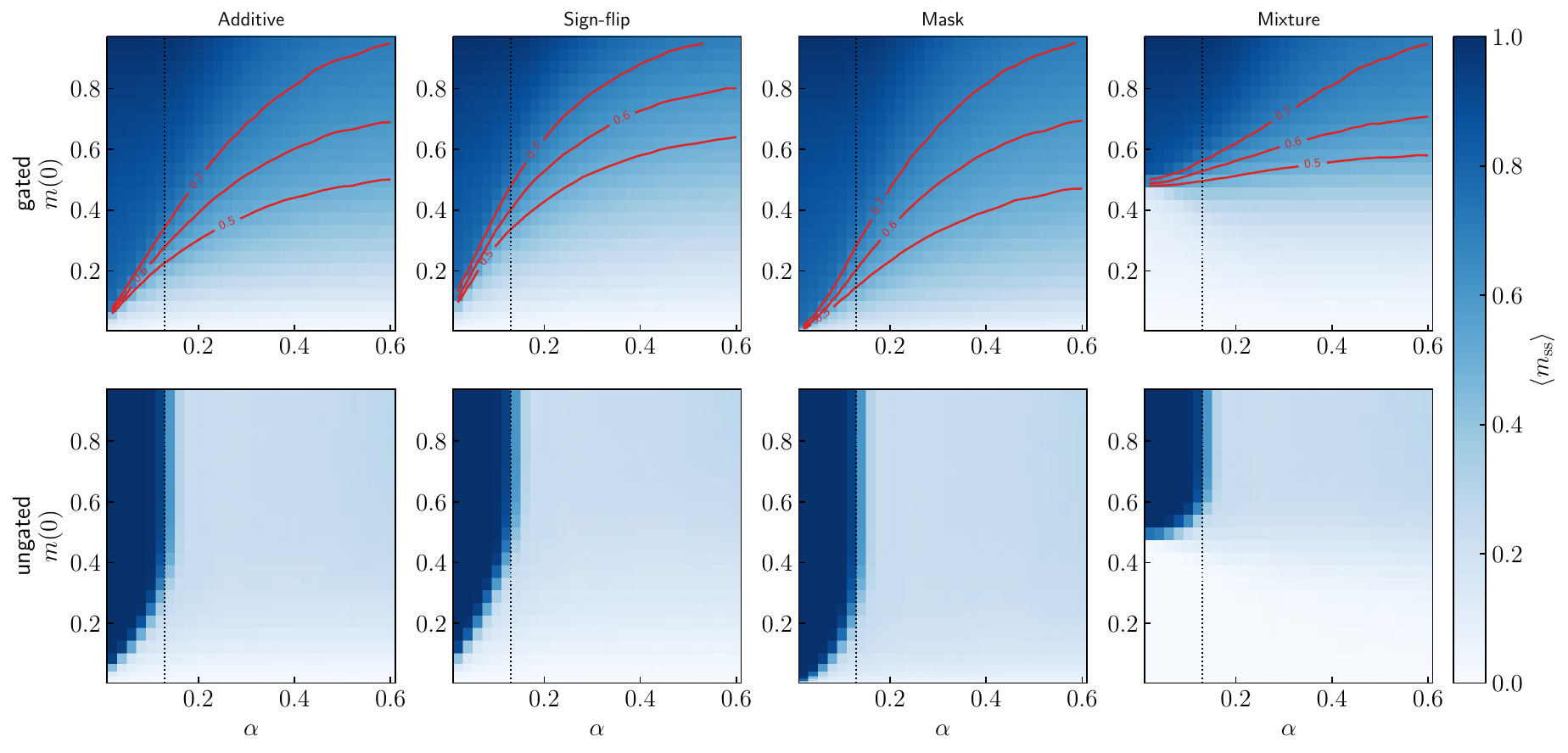}
    \caption{{\color{\editcolor}\textbf{Dependence of steady-state retrieval on how the initial condition is constructed.} Mean steady-state overlap $\langle m_\mathrm{ss}\rangle$ as a function of memory load $\alpha$ and initial overlap $m(0)$, for four constructions of the initial state $\mathbf{x}(0)$: additive (Gaussian noise added to the target pattern), sign-flip (a random fraction of components inverted), mask (a random fraction set to zero), and mixture (a second stored pattern superposed with amplitude $c_2 = 1$). Columns correspond to constructions; the top row is the gated network in the binary limit ($\gamma \to \infty$) and the bottom row the ungated network ($\gamma = 0$). Red curves are the level sets $\langle m_\mathrm{ss}\rangle = \theta$ for $\theta = \{ 0.5, 0.6, 0.7 \}$, that is, the retrieval boundaries $m_c(\alpha;\theta)$; inline labels give $\theta$. The dotted vertical line marks the ungated critical capacity $\alpha_c \simeq 0.13$. Note that the mixture cue retains $\mathcal{O}(1)$ overlap with a second stored pattern and is therefore not a corrupted version of the target. Many-body simulation parameters: $N = 1000, dt = 0.2$, averaged over 100 realizations of the quenched disorder $(\{\boldsymbol{\xi}^\mu\}, W_{ij}, \mathbf{z}(0))$ with 10 independent cues per realization at each value of $m(0)$.}
    }
    \label{fig:diff-initialization}
\end{figure*}
\begin{figure*}[!htbp]
    \centering
    \includegraphics[width=0.99\textwidth]{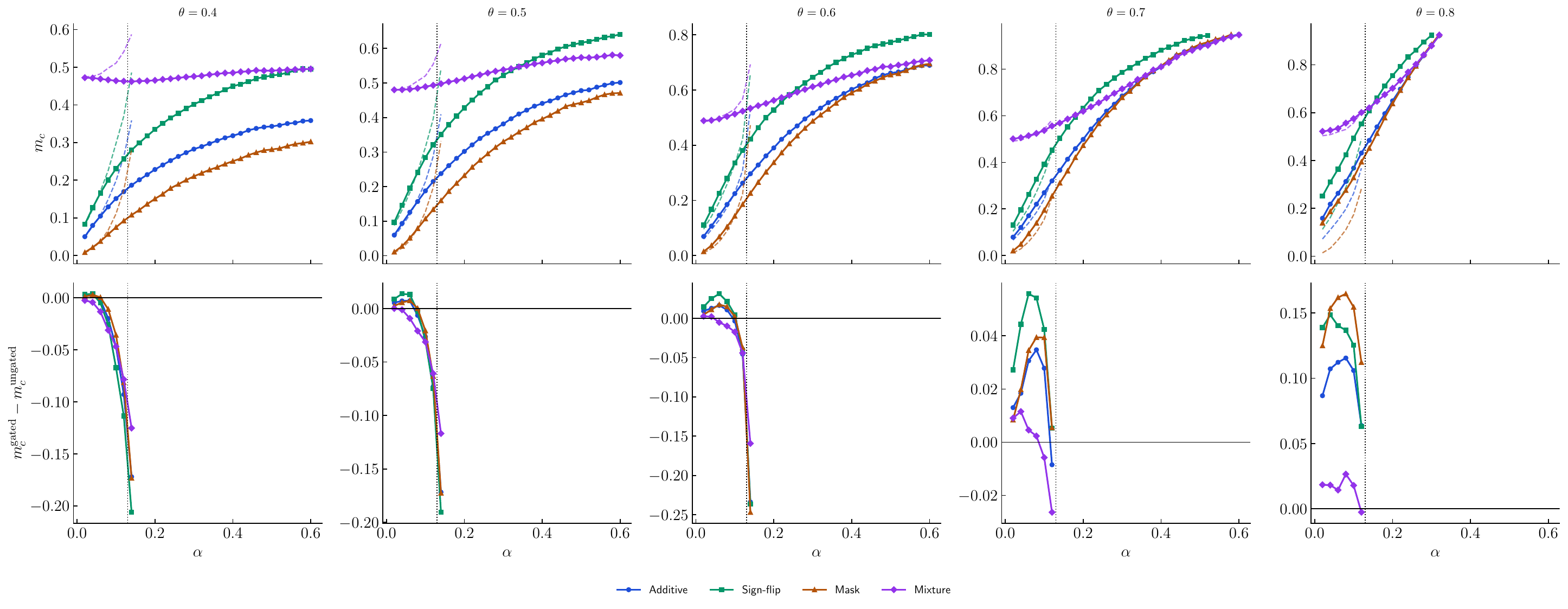}
    \caption{{\color{\editcolor}\textbf{Dependence of the retrieval boundary on the cue construction and on the choice of operational threshold.} (Top row) Retrieval boundary $m_c(\alpha;\theta)$, defined as the smallest initial overlap for which $\langle m_\mathrm{ss}\rangle > \theta$, plotted against memory load $\alpha$. Solid lines are gated networks in the binary limit $(\gamma \to \infty)$, dashed lines ungated $(\gamma = 0)$; colors and markers distinguish the four constructions of $\mathbf{x}(0)$  (additive, sign-flip, mask, mixture), as in Fig.~S2. A smaller $m_c$ corresponds to a larger empirical basin, since a weaker cue suffices. (Bottom row) Difference $m_c^{\gamma\to\infty} - m_c^{\gamma=0}$ at the same $\alpha$ and $\theta$; negative values indicate that the gated network retrieves from weaker cues, i.e. a larger basin. Each column corresponds to one threshold, $\theta = {0.4, 0.5, 0.6, 0.7, 0.8}$. Dotted vertical lines mark the ungated critical capacity $\alpha_c \simeq 0.13$; above it the ungated network has no retrieval boundary, so both the dashed curves and the differences terminate there. All $m_c$ values are obtained from the steady-state overlaps in Fig.~\ref{fig:diff-initialization}.}
    }
    \label{fig:diff-init_basins}
\end{figure*}

We determine whether the crossover effect and basin size of the gated network are robust to different constructions of the initial cue $\mathbf{x}(0)$. We test four methods — additive, sign-flip, mask, and mixture — of which the additive one was used for the main results. They are defined as follows:
\begin{align*}
    &\text{Additive: }\mathbf{x}(0) = c_1 \boldsymbol{\xi}^1 + \boldsymbol{\eta}, \\
    &\text{Sign-flip: }\mathbf{x}(0) = A \boldsymbol{v}_\mathrm{sf}, \\
    &\text{Mask: }\mathbf{x}(0) = A \boldsymbol{v}_\mathrm{m},  \\
    &\text{Mixture: }\mathbf{x}(0)  = c_1\boldsymbol{\xi}^1 + c_\mu\boldsymbol{\xi}^{\mu \neq 1}.
\end{align*}
Here $c_1$ is a scaling coefficient set by the target initial overlap $m(0)$, and $\boldsymbol{\eta} \sim \mathcal{N}(\mathbf{0}, I_N)$ is Gaussian noise. The amplitude $A$ sets the scale of the cue and caps the attainable initial overlap at $\tanh A$, so that high values of $m(0)$ remain reachable. The vector $\boldsymbol{v}_\mathrm{sf}$ is the pattern $\boldsymbol{\xi}^1$ with a fraction of its elements sign-inverted, that fraction being fixed by $m(0)$; likewise $\boldsymbol{v}_\mathrm{m}$ is $\boldsymbol{\xi}^1$ with a fraction of its elements masked (set to zero), again fixed by $m(0)$. Finally, $c_\mu$ scales the contribution of a second stored pattern $\boldsymbol{\xi}^{\mu \neq 1}$ to the mixture, drawn at random from the stored patterns other than the target. Throughout we set $A = 3$, so that $\tanh A \simeq 0.995$ and the full range of $m(0)$ is accessible, and $c_\mu = 1 $. For each initial cue type we perform many-body simulations of gated and ungated networks with $N = 1000$ and $dt = 0.2$, averaging over 100 instantiations of the quenched disorder $(\{\boldsymbol{\xi}^\mu\}, W_{ij}, \mathbf{z}(0))$  with 10 independent cues per realization at each value of $m(0)$.
\par
Figure~\ref{fig:diff-initialization} shows the mean steady-state overlap $\langle m_\mathrm{ss}\rangle$ as a function of memory load $\alpha$ and initial overlap $m(0)$, with one column per cue construction and gated and ungated networks in the top and bottom rows respectively. The gated networks bypass the sharp order-to-spin-glass transition for every construction, so the effect does not depend on how the cue is built; the ungated networks retain a critical capacity at $\alpha_c \simeq 0.13$ (dotted black line) in all four cases.
\par
The location of the retrieval boundary, by contrast, does depend on the construction. In the gated network the steady-state overlap varies continuously with $m(0)$ rather than switching between retrieval and failure, so no boundary exists without a convention: comparing basin sizes between gated and ungated networks requires an operational threshold $\theta$, with retrieval declared successful when $\langle m_\mathrm{ss}\rangle \geq \theta$. The threshold cannot be set close to unity, since the gated network loses fidelity even for $\alpha < \alpha_c$. The red contours in Fig.~\ref{fig:diff-initialization} are the level sets $\langle m_\mathrm{ss}\rangle = \theta$ for $\theta = \{0.5, 0.6, 0.7\}$, and their spread indicates how far the boundary moves with the threshold.
\par
Figure~\ref{fig:diff-init_basins} makes this comparison quantitative for all four cue constructions, using the same data as Fig.~\ref{fig:diff-initialization}. The top row shows the minimum initial overlap $m_c$ for which $\langle m_\mathrm{ss}\rangle \geq \theta$, for gated (solid lines with markers) and ungated (dashed lines) networks; the bottom row shows the difference $m_c^\mathrm{gated} - m_c^\mathrm{ungated}$, where negative values indicate a larger gated basin. Each column corresponds to a different threshold $\theta={0.4,0.5,0.6,0.7,0.8}$. At low $\theta$ the gated basin is the larger of the two; as $\theta$ increases the difference grows monotonically, crossing zero and becoming positive, so that at high values of $\theta$ the gated basin is smaller. Gating therefore broadens the range of cues from which some retrieval occurs, while requiring stronger cues for high-fidelity retrieval.
\par
These results show that whether gating preserves the basin of attraction depends on the operational threshold chosen. At low fidelity thresholds the gated basin is the larger of the two, whereas at high thresholds reaching a given overlap requires a stronger cue than in the ungated network. Gating therefore broadens the range of cues from which some retrieval occurs, while narrowing the range from which high-fidelity retrieval occurs. We note that this comparison arises only below the ungated critical capacity, because for $\alpha > \alpha_c$ the ungated network has no retrieval boundary at any threshold; in that regime the gated network retrieves where the ungated network does not, at every $\theta$. The finite-size analysis that follows shows, however, that the comparison shifts in favor of the gated network as $N$ grows at every threshold examined, so the threshold dependence reported here is weaker in the thermodynamic limit than at the system sizes used above.

\begin{figure*}[!htbp]
    \centering
    \includegraphics[width=0.94\textwidth]{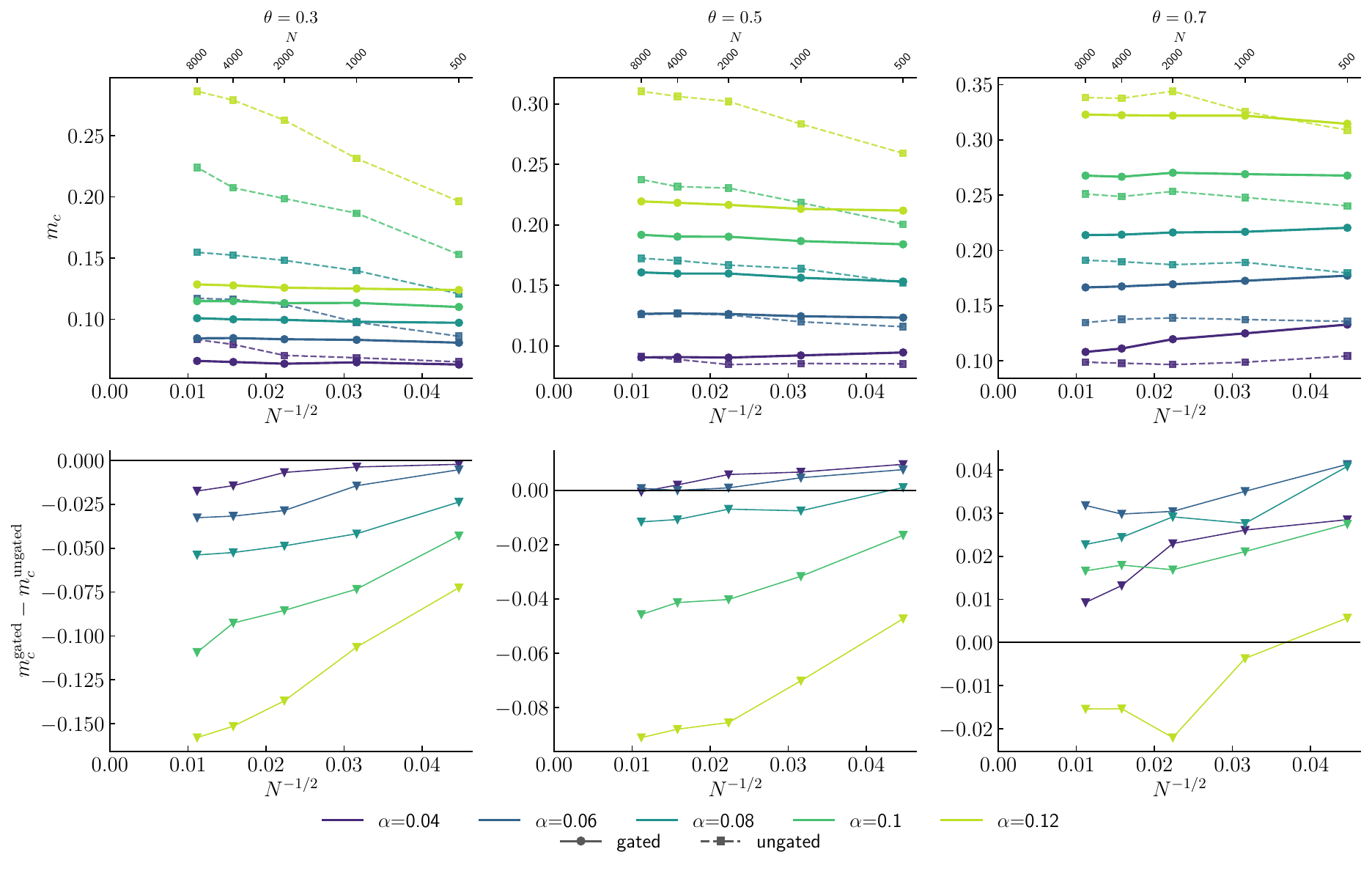}
    \caption{{\color{\editcolor}\textbf{Finite-size behavior of the retrieval boundary $m_c$.} (Top) Retrieval boundary $m_c(\alpha;\theta)$ against $N^{-1/2}$, where $N$ is the network size, so that the thermodynamic limit lies at the left edge; the upper axis gives the corresponding $N$. Each column corresponds to a different $\theta = \{ 0.3,0.5,0.7\}$. Circles with solid lines are gated networks ($\gamma\to\infty$), squares with dashed lines ungated ($\gamma=0 $); color denotes memory load. (Bottom) The difference $m_c^{\rm gated}-m_c^{\rm ungated}$, with negative values indicating that the gated network retrieves from weaker cues, i.e. a larger empirical basin. Many-body simulations use 60 disorder realizations with 10 independent cues each per value of $m(0)$.}
    }
    \label{fig:finite-size-basin-effs}
\end{figure*}

\begin{figure*}[!htbp]
    \centering
    \includegraphics[width=0.7\textwidth]{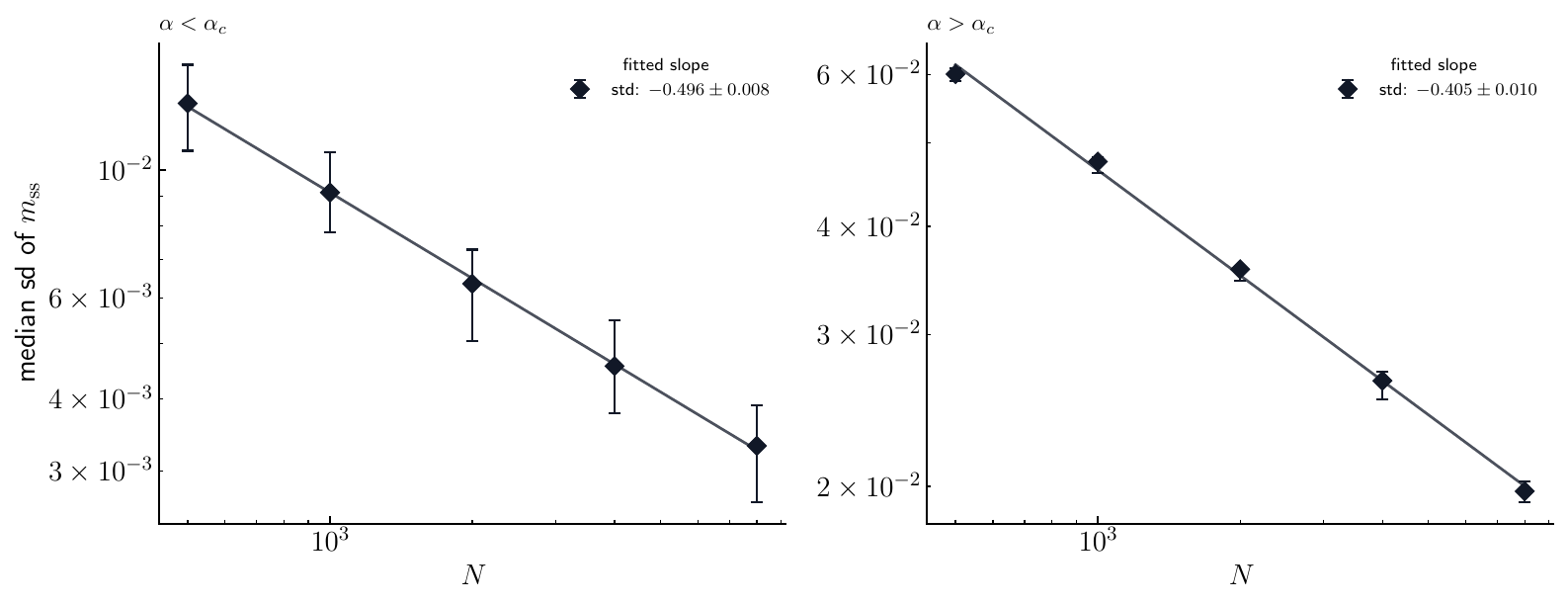}
    \caption{{\color{\editcolor}\textbf{Standard deviation of the steady-state overlap versus system size.} Median standard deviation of $m_\mathrm{ss}$ across disorder realizations and cues, for gated networks with $N = \{500, 1000, 2000, 4000, 8000\}$, below the critical memory capacity ($\alpha < \alpha_c$, left) and above it ($\alpha > \alpha_c$, right). Medians are over all sampled initial overlaps, $0.3 \le m(0) \le 0.9 $. Error bars are $68\%$  bootstrap intervals; solid lines are least-squares fits with the exponent given in each legend.}
    }
    \label{fig:finite-size-effs}
\end{figure*}

\subsection{Finite size effects: Basin size and $m_\mathrm{ss}$}
\label{sec:finite-size}

In this section we show how empirical basin sizes for gated and ungated systems scale with network size and that the mean $\langle m_\mathrm{ss}\rangle$ is a well-defined statistical quantity despite the multistability and manifold of fixed points found in gated networks. 
\par
Figure~\ref{fig:finite-size-basin-effs} (top) shows that the gated retrieval boundary $m_c^\mathrm{gated}$ (circles) is essentially independent of system size at every threshold examined, whereas the ungated boundary $m_c^\mathrm{ungated}$ (squares) increases with $N$. The gated empirical basin therefore remains fixed as the thermodynamic limit is approached while the ungated basin shrinks, so that the difference between the two (bottom, triangles) becomes increasingly negative with increasing $N$. The finite-size trend therefore favors the gated network at every threshold examined, and our results at $N = 1000$ understate rather than overstate its basin advantage.
\par
Meanwhile, each point in the phase diagrams of Fig.~\ref{fig:main} is an average over many disorder realizations and cues; if the spread of $m_\mathrm{ss}$ at fixed $(\alpha, m(0))$ persisted in the thermodynamic limit, that average would not characterize any individual network and the diagrams would describe only the ensemble.
\par
We therefore measure how that spread scales with system size. For each $N = \{500, 1000, 2000, 4000, 8000\}$ we simulate gated networks over a reduced grid of memory loads spanning both sides of the critical capacity, with $100$ realizations of the quenched disorder and $10$ independent cues per realization at each initial overlap, and compute the standard deviation of $m_\mathrm{ss}$ for each $(\alpha, m(0))$ cell.
\par
Figure~\ref{fig:finite-size-effs} shows the result. In both regimes, $\alpha < \alpha_c$ (left) and $\alpha > \alpha_c$ (right), the width decreases as a power of $N$, with fitted exponents between $-0.4$ and $-0.5$; the exponents vary by at most $0.04$ across the range of initial overlaps averaged over. The spread therefore vanishes as $N \to \infty: m_\mathrm{ss}$ is self-averaging, and becomes a deterministic function of $\alpha$ and $m(0)$ in the thermodynamic limit. The multistability reported in Sec.\ts\ref{sec:multistability} is thus a continuum of steady-state overlaps indexed by the initial overlap, and not a residual scatter about a mean.
\color{black}

\section{Relevant literature}
In this section, we briefly review related literature, with the goal of clarifying how these formulations and operating regimes differ from our self-adaptively gated associative memory network.

\subsection{Krotov's hierarchical networks}
Here we briefly review the multi-layer formulation by Krotov~\cite{krotov2021hierarchicalassociativememory} (and related Krotov--Hopfield constructions~\cite{Krotov-Hopfield21}) to highlight differences from our self-adaptively gated associative memory model.
\par
Krotov and Hopfield proposed classes of two-layer networks as a biologically plausible realization of dense associative
memories using only pairwise interactions between two sets of variables (see also \cite{Kozachkov-Slotine-Krotov25}). 
They consider state variables
$\mathbf{x}\in\mathbb{R}^{N}$ and $\mathbf{h}\in\mathbb{R}^{P}$, coupled through an inter-layer interaction matrix
$\bm{\Xi}\in\mathbb{R}^{N\times P}$ with columns $\bm{\xi}^\mu\in\mathbb{R}^{N}$, which represent the memorized patterns.
A canonical continuous-time neuron--substrate (or feature--hidden) dynamics can be written as
\begin{align}
    \tau_x\partial_t x_i &= -x_i + \sum_{\mu=1}^{P}\xi_{i}^{\mu} f_\mu(\mathbf{h}) \ ,
    \label{eq:krotov_x}\\
    \tau_h\partial_t h_\mu &= -h_\mu + \sum_{i=1}^{N}\xi^{\mu}_{i} g_i(\mathbf{x}) \ ,
    \label{eq:krotov_h}
\end{align}
where the outputs can be expressed as variations of layer-wise potentials, \textit{e.g.}, 
$f_\mu(\mathbf{h})=\partial\mathcal{L}_h(\mathbf{h})/\partial h_\mu$ and 
$g_i(\mathbf{x})=\partial\mathcal{L}_x(\mathbf{x})/\partial x_i$ for a suitable choice of $\mathcal{L}_h,\mathcal{L}_x$.
The crucial structural feature is that the two layers share the same inter-layer interaction matrix 
$\bm{\Xi}=(\bm{\xi}^1,\ldots,\bm{\xi}^P)$: the memories $\{\bm{\xi}^\mu\}$ appear in both the $\mathbf{x}\to\mathbf{h}$ 
and $\mathbf{h}\to\mathbf{x}$ pathways. 
For appropriate symmetry/consistency conditions and compatible choices of nonlinearities, this variational structure admits the existence of a Lyapunov (energy) function.
Assuming an adiabatic timescale separation $\tau_x\gg\tau_h$, the second equation is approximated (quasi-)stationary as 
$\mathbf{h}\simeq{\bm{\Xi}}^\top\mathbf{g}(\mathbf{x})$.
Substituting this into the first equation gives an effective single-layer dynamics
\begin{equation}
    \tau_x\partial_t x_i \simeq -x_i + \sum_{\mu=1}^{P}\Xi_{i\mu}\,
    f_\mu\!\big(\bm{\Xi}^\top\mathbf{g}(\mathbf{x})\big) \ .
    \label{eq:krotov_reduced}
\end{equation}
Depending on the choice of nonlinearities encoded by $\mathcal{L}_h$ and $\mathcal{L}_x$ (\textit{e.g.}, polynomial-type or exponential-type choices), 
the second term in Eq.~\eqref{eq:krotov_reduced} produces effective higher-order interactions and recovers dense associative memory energy functions after integrating out hidden variables.
\par
A key contrast with our gated associative memory model is that, in Krotov-type two-layer models, the objects that encode memories/keys/values are embedded in the shared inter-layer matrix $\bm{\Xi}$ that couples $\mathbf{x}$ and $\mathbf{h}$; 
the two layers therefore inherently share the same memory-bearing structure. 
In our model, by construction, the stored patterns are encoded only in the neuronal synaptic couplings ($J$), whereas the gating degrees of freedom and their couplings ($W$) act as modulators and do not share the same pattern structure as $J$. 
\par
Additionally, Krotov-type models are typically formulated with an explicit Lyapunov/energy structure (under appropriate symmetry conditions), yielding an effectively equilibrium-like attractor relaxation picture. 
Our gated dynamics can generically violate detailed balance and need not admit a time-independent energy function; the effective landscape is state-dependent through gating.

\subsection{Masse, Grant, \& Freedman: preventing catastrophic forgetting during sequential learning}
Masse \textit{et al.}~\cite{Masse-Grant-Freedman18} study catastrophic forgetting as a \emph{training-time} phenomenon in continual learning: sequential gradient updates overwrite previously useful weights, degrading performance on earlier tasks. 
They introduce \emph{context-dependent gating} (XdG), where an externally provided task/context cue gates activity so each task uses a sparse, partially nonoverlapping subnetwork.
In contrast, our work does not address sequential task learning or weight overwriting.
Assuming that synapses are fixed after Hebbian learning, we study \emph{retrieval-time} catastrophic breakdown of associative recall at high load (Hopfield-style collapse/spin-glass transition) and show that self-adaptive, activity-dependent gating acts during retrieval to reorganize attractors and sustain robust recall beyond the standard critical capacity.

\subsection{Podlaski, Agnes, \& Vogels: context-modular associative memory via neuronal/synaptic gating}
Podlaski \textit{et al.}~\cite{Podlaski-Agnes-Vogels25} start from a \emph{context-modular} associative memory in which the effective interaction structure is \emph{context-dependent by construction}: 
activating a particular context imposes a corresponding background network state that selects a context-specific set of active neurons and/or synapses (neuronal vs. synaptic gating), thereby defining a context-dependent effective connectivity. 
In our model, by contrast, the starting synaptic memory couplings are not pre-partitioned into externally indexed contexts; 
instead, self-adaptive gating emerges as an internal dynamical variable and modifies retrieval \emph{state-dependently} during the recall process. 
Their \emph{neuronal gating} is conceptually closest to our mechanism in the sense that it modulates which neuronal degrees of freedom participate in retrieval; 
however, their gates are \emph{context-dependent} (set by the active context cue) rather than \emph{state-/activity-dependent} variables evolving with the ongoing retrieval dynamics. 
While they develop their theory in a discrete-state setting, the main conceptual distinction from our work is the \emph{source} of gating (externally selected context versus internally generated, activity-dependent dynamics). 
As discussed in the main text, our time-constant-modulating gate can be motivated axiomatically from quasi-invariance to nonlinear time reparametrizations (“time warps”) of neural activity~\cite{Tallec-Ollivier18}, suggesting our model as a minimal continuous-state realization of this generic gating principle.

{\color{\editcolor}
\subsection{Fiete and colleagues: fixed memory scaffolds for robust addressing}
A structurally different route to robust, high-capacity associative memory builds error correction into the internal representation.
Chaudhuri and Fiete introduced a sparse bipartite expander network that supports exponentially many stable states with basins capable of correcting errors in a finite fraction of the nodes through the network's own dynamics \cite{Chaudhuri-Fiete19}. 
MESH separates this internal stabilization from content storage: a predetermined set of content-independent scaffold states is heteroassociated with arbitrary external patterns, replacing the memory cliff with a graceful trade-off between the number of patterns and the information recovered per pattern \cite{Sharma-Chandra-Fiete22}. 
Vector-HaSH gives this principle an explicit biological realization in a neocortical--entorhinal--hippocampal architecture whose scaffold is
constructed from grid-cell states \cite{Chandra-Sharma-Chaudhuri-Fiete25,Sreenivasan-Fiete11}. 
Thus, the expander network is primarily motivated by coding theory, whereas MESH and Vector-HaSH additionally draw on the organization of biological memory circuits.

}

\renewcommand{\bibsection}
  {\section*{Supplementary References}}

\putbib[ref_hopfield]

\end{bibunit}
\endgroup

\end{document}